\documentclass[manuscript]{aastex}
\usepackage{color}

\shorttitle{SASI}
\shortauthors{Iwakami et al.}

\begin{document}

\title{Parametric Study of Flow Patterns behind the Standing Accretion Shock Wave for Core-Collapse Supernovae}

\author{Wakana Iwakami}
\affil{Yukawa Institute for Theoretical Physics, Kyoto University, Oiwake-cho, Kitashirakawa, Sakyo-ku, Kyoto, 606-8502, Japan}
\affil{Advanced Research Institute for Science and Engineering, Waseda University, 3-4-1, Okubo, Shinjuku, Tokyo, 169-8555, Japan}

\author{Hiroki Nagakura}
\affil{Yukawa Institute for Theoretical Physics, Kyoto University, Oiwake-cho, Kitashirakawa, Sakyo-ku, Kyoto, 606-8502, Japan}

\and

\author{Shoichi Yamada}
\affil{Advanced Research Institute for Science and Engineering, Waseda University, 3-4-1, Okubo, Shinjuku, Tokyo, 169-8555, Japan}
\email{wakana@heap.phys.waseda.ac.jp}

\begin{abstract}
In this study, we conduct three-dimensional hydrodynamic simulations systematically to investigate the flow patterns behind the accretion shock waves that are commonly formed in the post-bounce phase of core-collapse supernovae.
Adding small perturbations to spherically symmetric, steady, shocked accretion flows, we compute the subsequent evolutions to find what flow pattern emerges as a consequence of hydrodynamical instabilities such as convection and standing accretion shock instability (SASI) for different neutrino luminosities and mass accretion rates.
Depending on these two controlling parameters, various flow patterns are indeed realized.
We classify them into three basic patterns and two 
intermediate
ones; the former includes sloshing motion (SL), spiral motion (SP) and multiple buoyant bubble formation (BB); the latter consists of spiral motion with buoyant-bubble formation (SPB) and spiral motion with pulsationally changing rotational velocities (SPP).
Although the post-shock flow is highly chaotic, there is a clear trend in the pattern realization.
The sloshing and spiral motions tend to be dominant 
for high accretion rates and low neutrino luminosities, and
multiple buoyant bubbles prevail for low accretion rates and high neutrino luminosities.
It is interesting that the dominant pattern is not always identical between the semi-nonlinear and nonlinear phases near the critical luminosity; the 
intermediate
cases are realized in the latter case. 
Running several simulations with different random perturbations, we confirm that the realization of flow pattern is robust in most cases.
\end{abstract}


\section{INTRODUCTION \label{sec_intro}}

Core-collapse supernova has been studied for more than a half century \citep[see][for latest reviews]{janka12, burrows13}.
A number of one-dimensional computational simulations, where spherical symmetry is assumed, have consistently demonstrated that a shock wave generated by gravitational core-collapse can not reach the surface of progenitor \citep[e.g.,][]{liebend05,sumiyoshi05}, except for the low-mass progenitors \citep{kitaura06}.
Moreover, some observations have indicated that core-collapse supernova is intrinsically asymmetric \citep[e.g.,][]{wang01, leonard06}.
It is hence believed that multi-dimensionality is essentially important for successful explosion.
There are a couple of possible multi-dimensional effects: rotation, magnetic fields, hydrodynamical instabilities, and so on.
Among other things, we pay attention to the hydrodynamical instabilities in this paper.

At present, two instabilities are attracting researchers' interest: neutrino-driven convection and standing accretion shock instability (SASI).
The neutrino-driven convection is induced by the negative entropy gradient that arises naturally in the gain layer due to neutrino heating \citep[e.g.][]{bethe90}.
A lot of two-dimensional (2D) \citep[e.g.,][]{herant92, burrows92, janka96} and three-dimensional (3D) simulations \citep[e.g.,][]{fryer02,nordhaus10,couch13a,dolence13, murphy13, ott13} have demonstrated its importance.
On the other hand, SASI is an instability associated with shock deformations \citep[e.g.][]{foglizzo07}.
Its notable feature is the dominance of low $l$ modes ($l=1$ or 2 in particular), where $l$ denotes the polar index of spherical harmonics.
A number of studies on SASI have been done by
2D \citep[e.g.,][]{blondin03, ohnishi06, burrows06, scheck08},
3D simulations \citep[e.g.,][]{blondin07, iwakami08, fernandez10, burrows12, takiwaki12, wongwathanarat13, hanke12, hanke13}, 
linear analyses \citep[e.g.,][]{foglizzo07, yamasaki07, yamasaki08, foglizzo09, fernandez09b, guilet10, guilet12},
and SWASI experiments \citep{foglizzo12}.
It is now becoming a consensus that SASI is driven by the advective-acoustic cycle.

These results have shown that 3D numerical simulations give the different flow patterns depending on the group;
strong dipolar-oscillations or rotations are observed by \citet{blondin07} and \citet{hanke13}, but high-entropy bubbles moving upward are found instead by \citet{fryer02}, \citet{iwakami08}, \citet{nordhaus10}, and \citet{dolence13}.
More importantly, 
\citet{burrows12} and \citet{hanke13} suggest that the emergence of particular flow patterns is related to the neutrino luminosity and mass accretion rate.
No systematic study on this issue has been done so far, however, and it is the subject of this paper.
The flow pattern is important, since times related to some observable such as the eventual morphology of explosion, pulsar kick and spin, and neutrino signal and gravitational wave.
The study of flow patterns might be also useful for the understanding of the mechanism of convection and SASI, one of the key factors to determine the critical neutrino luminosity for explosion.

In order to address the problem described above,
i.e., which flow pattern appears under what condition,
we perform experimental 3D simulations of the accretion flows through a standing shock wave, which represent approximately the post-bounce phase of supernova cores.
For that purpose, we extend our previous studies \citep[][]{ohnishi06, iwakami08, iwakami09a, iwakami09b}.
In this study, the mass accretion rate and neutrino luminosity are assumed to be constant in time, and the unperturbed flows are spherically symmetric and steady initially.
We employ the so-called light-bulb approximation for neutrino heating/cooling.
We consider 9 combinations of a neutrino luminosity and mass accretion rate.
For each combination, we run several simulations, changing a random perturbation imposed at the beginning of computation.
This is necessary because the post-shock flow is turbulent and highly stochastic and we cannot draw a firm conclusion from a single realization.

The organization of this paper is as follows.
We describe models and numerical methods in Section 2,
present the main results in Section 3,
and conclude the paper in Section 4.

\section{MODELS AND FORMULATIONS \label{sec_models}}

\subsection{Numerical Setups \label{sec_setup}}
The numerical method is the same as that used in our previous paper \citep{iwakami08}.
The ZEUS-MP/2 code \citep{hayes06} is modified to solve the accretion flow irradiated by neutrinos emitted from the proto-neutron star (PNS); the tabulated EOS by \citet{shen98} is implemented in the code;
the light bulb approximation \cite[][for details]{ohnishi06} is adopted for the neutrino heating and cooling.
It is assumed that neutrinos are emitted isotropically from PNS with prescribed fluxes and the matter outside PNS is optically thin.
Only electron-type neutrinos and anti-neutrinos are considered in this study.
Their temperatures are also assumed to be constant in time and set to the typical values in the post-bounce phase, $T_{\nu_{\mathrm{e}}} = 4 $~MeV and $T_{\bar{\nu}_{\mathrm{e}}} = 5 $~MeV.
The mass of PNS is fixed to $M_{\rm PNS} =1.4 M_\odot$.
As model parameters, the neutrino luminosity and mass accretion rate are varied in the range of $L_\nu = 2.0 - 6.0 \times 10^{52}$~erg~s$^{-1}$ and $\dot{M} = 0.2 - 1.0~M_{\odot}$~s$^{-1}$, respectively.
All models presented in this paper are summarized in Table~\ref{tbl_param}.
More than
three different realizations are computed for each model.

A staggered grid is used in the ZEUS-MP/2.
The spherical polar coordinates $(r, \theta, \phi)$ are adopted. 
The computational domain covers the entire solid angle.
The radial inner boundary is located at $r_{\rm in} \sim 28 - 50$km,
and the outer boundary at $r_{\rm out} \sim 500 - 1000$km.
The computational region is divided into $300\times30\times60$ grid cells in $r\times\theta\times\phi$ directions, as explained below.
The radial size $\Delta r_{i}$ and position $r_i$ of the $i$-th cell are given as
\begin{equation}
\Delta r_{i} = r_c \Delta r_{i-1},\ \ \ r_i = \Delta r_i/(r_c -1),\ \ \ (i=1, 2, \cdot\cdot\cdot, N),
\end{equation}
where $N$ denotes the total number of grid cells, and $r_c$ the common ratio of the geometric sequence,
which $r_c$ is fixed to be 1.01, that is, the radial resolution is 1\% of the radius. 
The inner boundary of the computational domain is located at $\rho = 10^{11}$ g cm$^{-3}$ in the spherically symmetric, steady, unperturbed states.
It roughly coincides with the neutrino sphere and hence depends on $L_\nu$ and $\dot{M}$.
The locations of outer boundary $r_\mathrm{out}$ are also different among models,
since $N$ is fixed to 300.
The values of $r_{\rm in}$ and $r_{\rm out}$ in each model are listed in Table~\ref{tbl_param}. 

The boundary conditions for other quantities are set as follows.
The velocity, density, internal energy, and electron fraction are fixed to their initial values at the outer boundary.
At the inner boundary, on the other hand,
the density, internal energy, and electron fraction are assumed to have a vanishing gradient in the direction normal to the inner boundary, 
and the radial velocity is fixed to the initial value,
whereas $\theta$ and $\phi$ components are set so that the conservation of angular momentum should be conserved.

\begin{figure}
\epsscale{0.90}
\plotone{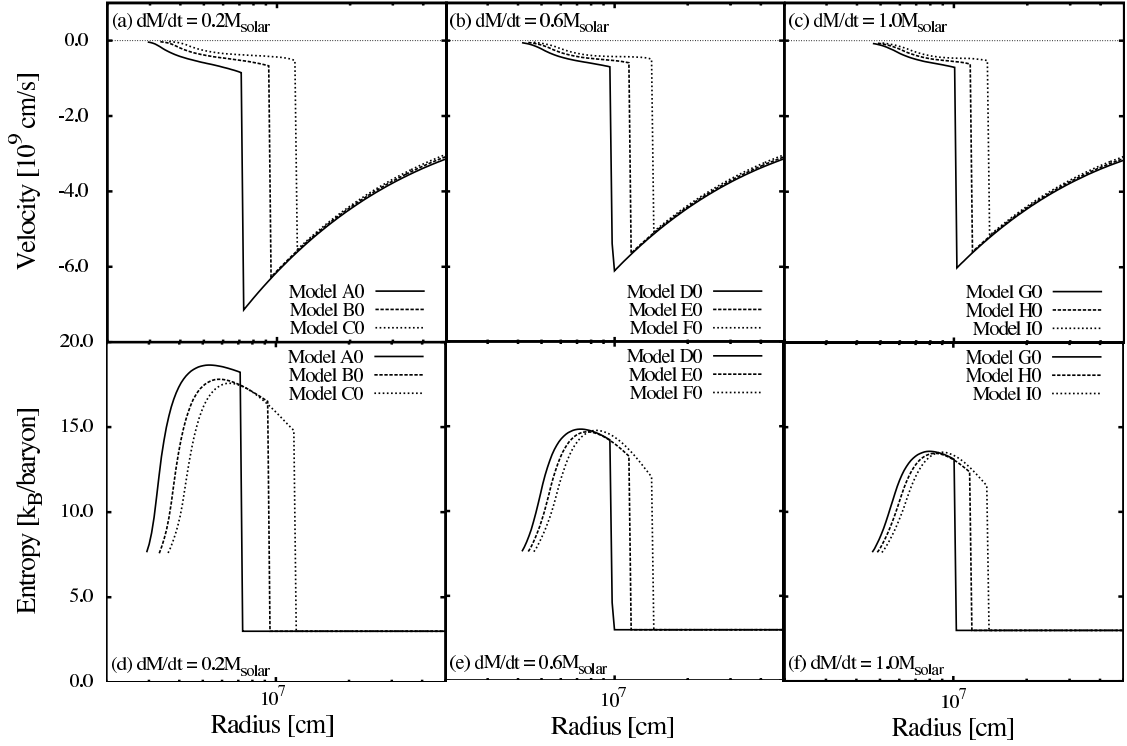}
\caption{
The distributions of the radial velocity and entropy in the unperturbed, spherically symmetric, steady accretion flows.
The former is shown in the upper panels, and the latter in the lower panels.
The left, middle, and right panels are the results for $\dot{M} = 0.2 M_\odot$, $0.6 M_\odot$, and $1.0 M_\odot$, respectively.
\label{fig_init}}
\end{figure}

The initial conditions, or spherically symmetric, steady, shocked accretion flows, are obtained in the same manner as \citet{yamasaki06}, except that we do not take into account self-gravity in and determine the location of the inner boundary not based on the optical depth but on the density for simplicity.
The profiles of radial velocity and entropy in the initial unperturbed states are displayed in Figure~\ref{fig_init}.
As observed in the upper panels, the standing shock wave decelerates the inward flow abruptly and the shocked matter slows down further as it approaches the proto neutron star (PNS).
As shown in the lower panels, on the other hand, all the models considered in this paper have a region with a negative entropy gradient, which corresponds to the gain region as is well known.
The combinations of $L_\nu$ and $\dot{M}$ adopted in this paper hence give the models that are unstable to convection in the classical sense. Discussions on this issue based on the so-called $\chi$ parameter \citep{foglizzo06} will be given later.
It will be also useful to mention that the parameter combinations correspond to those that give the 
oscillatory,
unstable modes in the linear analysis by \citet{yamasaki07}, which the authors interpreted as 
SASI.
Furthermore, we confirm that our models are stable to radial perturbations, the fact also consistent with the results of \citet{yamasaki07}.
In order to induce non-spherical instabilities, we add cell-wise, random perturbations within 1\% to the radial velocity at the beginning of computation.

\subsection{Mode Analysis \label{sec_mode}}
We employ mode analysis to judge which flow pattern is dominant.
The deformation of shock surface can be expanded as a linear combination of the spherical harmonic components,
\begin{equation}
R_\mathrm{sh}(\theta, \phi, t) = \sum^{\infty}_{l=0} \sum^{l}_{m=-l} c^m_{l} (t) \, Y^m_{l}(\theta, \phi),
\label{eq_sph}
\end{equation}
where $Y^m_{l}$ is expressed by the associated Legendre polynomial $P^m_{l}$ as
\begin{equation}
Y^m_{l} = K^m_{l} P^m_{l}(\cos \theta) \, e^{im\phi},
\ \ \ K^m_{l} = \sqrt{\frac{2l+1}{4\pi}\frac{(l-m)!}{(l+m)!}}.
\label{eq_y}
\end{equation}
The expansion coefficients can be obtained as follows,
\begin{equation}
c^m_{l} (t) =\int^{2\pi}_0 \! \! \! \! d\phi  \! \int^{\pi}_0 \! \! d\theta \, \sin \theta \, R_\mathrm{sh}(\theta, \phi, t) \, Y^{m*}_{l} (\theta, \phi),
\label{eq_c}
\end{equation}
where the superscript * denotes complex conjugation.

We define the following quantities:
\begin{equation}
A_l(t) = \sqrt{\Sigma^l_{m=-l} |c^m_l(t)/c^0_0(t)|^2},
\label{eq_A}
\end{equation}
\begin{equation}
A_{1, 2}(t) =  \sqrt{\Sigma^2_{l=1}\Sigma^l_{m=-l} |c^m_l(t)/c^0_0(t)|^2}, 
\label{eq_A12}
\end{equation}
\begin{equation}
A_{4, 5} (t) =  \sqrt{\Sigma^5_{l=4}\Sigma^l_{m=-l} |c^m_l(t)/c^0_0(t)|^2},
\label{eq_A45}
\end{equation}
and utilize also their time-averages:
\begin{equation}
\bar{A}_l = \frac{1}{T}\int^{t_e}_{t_s} A_l(t) dt,
\label{eq_Ab}
\end{equation}
\begin{equation}
\bar{A}_{1,2} = \frac{1}{T}\int^{t_e}_{t_s} A_{1,2}(t) dt,
\label{eq_A12b}
\end{equation}
\begin{equation}
\bar{A}_{4,5} = \frac{1}{T}\int^{t_e}_{t_s} A_{4,5}(t) dt,
\label{eq_A45b}
\end{equation}
where $T=t_e-t_s$ is the integral time, $t_s$ is the starting time, and $t_e$ is the ending time for integration.

\subsection{$\chi$ Parameter \label{sec_chi}}
The classical condition of convective instability is the existence of the negative entropy gradient.
In the presence of matter flow, however, this condition is not sufficient.
In fact, matter passes through a convectively unstable region in a finite time in general.
If this advection time is shorter than the growth time of convection, then the instability is suppressed. 
\citet{foglizzo06} propose a new criterion that takes the advection of matter into account.
They introduce the $\chi$ parameter, which is the ratio of the advection time to the local timescale of buoyancy and is defined as
\begin{equation}
\chi \equiv\int_{r_{\rm gain}}^{r_{\rm sh}} \left|\frac{N}{u_r}\right| dr,
\label{eq_chi}
\end{equation}
where
the integration is done over the gain region with $r_{\rm gain}$ being the gain radius, i.e. the boundary between cooling and heating region, and $r_{\rm sh}$ being the shock radius.
$u_r$ is the radial velocity.
The Brunt-V\"{a}is\"{a}l\"{a} frequency $N$ is defined as
\begin{equation}
N^2\equiv\left[
 \frac{1}{p}\left(\frac{\partial p}{\partial   S}\right)_{\rho,Y_e}\frac{dS  }{dr}
+\frac{1}{p}\left(\frac{\partial p}{\partial Y_e}\right)_{\rho,  S}\frac{dY_e}{dr}
\right]
\frac{g}{\Gamma_1},
\label{eq_norg}
\end{equation}
\begin{equation}
\Gamma_1 = \left(\frac{\partial \ln{p}}{\partial \ln{\rho}}\right)_{S, Y_e},
\label{eq_gamma}
\end{equation}
where $p$, $\rho$, $S$, $Y_e$, and $G$ are the pressure, density, entropy, electron fraction, and gravitational constant, respectively.
$g$ is the gravitational acceleration and given approximately in the gain region as $g = \frac{GM_{\rm PNS}}{r^2}$, in which $M_{\rm PNS}$ is the PNS mass.
Substituting Eq.(\ref{eq_gamma}) into Eq.(\ref{eq_norg}),
we obtain the following expression,
\begin{equation}
N^2 =\left|\frac{1}{\Gamma_1 p}\frac{dp}{dr}-\frac{1}{\rho}\frac{d\rho}{dr}\right|g,
\label{eq_n}
\end{equation}
which is actually used for the calculation of $\chi$ in the following.

Obtained originally in the linear analysis \citep{foglizzo06}, the above formula is meant for the application to 1D steady states.
In the literature, however, it has been also applied to somehow defined mean flows in the full-fledged turbulent induced by convection \citep[e.g.,][]{fernandez13, couch13b}.
Following these practices, we define for non-exploding models in this paper the mean flow by time-averaging the quantities in the flow from the onset of the quasi-steady phase up to the end of computation:
\begin{equation}
\bar{q}(r,\theta,\phi)=\frac{1}{T}\int^{t_e}_{t_s} q(r,\theta,\phi,t) dt,
\label{eq_tb}
\end{equation}
where $q(r,\theta,\phi,t)$ means an arbitrary quantity (i.e., $\rho$, $u_r$, and so on).
In applying the above criterion, we further take an angle-average over the entire solid angle at each radius, 
\begin{equation}
\bar{q}_{\rm 1D}(r) = \frac{1}{4\pi}\int_0^{4\pi} \bar{q}(r,\theta,\phi) d\Omega.
\label{eq_ab}
\end{equation}
The parameter $\bar{\chi}_{\rm 1D}$ is then obtained as
\begin{equation}
\bar{\chi}_{\rm 1D} = \int_{\bar{r}_{\rm gain 1D}}^{\bar{r}_{\rm sh 1D}} \left|\frac{\bar{N}_{\rm 1D}(r)}{\bar{u}_{r {\rm 1D}}(r)}\right| dr,
\label{eq_kaib}
\end{equation}
\begin{equation}
\bar{N}^2_{\rm 1D} =\left|\frac{1}{\bar{\Gamma}_{1\rm 1D} \bar{p}_{\rm 1D}}\frac{d\bar{p}_{\rm 1D}}{dr}-\frac{1}{\bar{\rho}_{\rm 1D}}\frac{d\bar{\rho}_{\rm 1D}}{dr}\right|g,
\label{eq_nb}
\end{equation}
where the quantities with a bar are calculated for the angle-averaged mean flow.
The thermodynamical variables, $\bar{p}_{\rm 1D}$ and $\bar{\Gamma}_{1\rm 1D}$, are given by the EOS table as a function of $\bar{\rho}_{\rm 1D}$, $\bar{e}_{\rm 1D}$ and $\bar{Y}_{e\rm 1D}$.

\section{RESULTS AND DISCUSSIONS \label{sec_result}}

\subsection{Overview of the Results \label{sec_over}}

In this section, we describe the basic characteristics of each model presented in Table~\ref{tbl_kai}.

\begin{figure}
\epsscale{.45}
\plotone{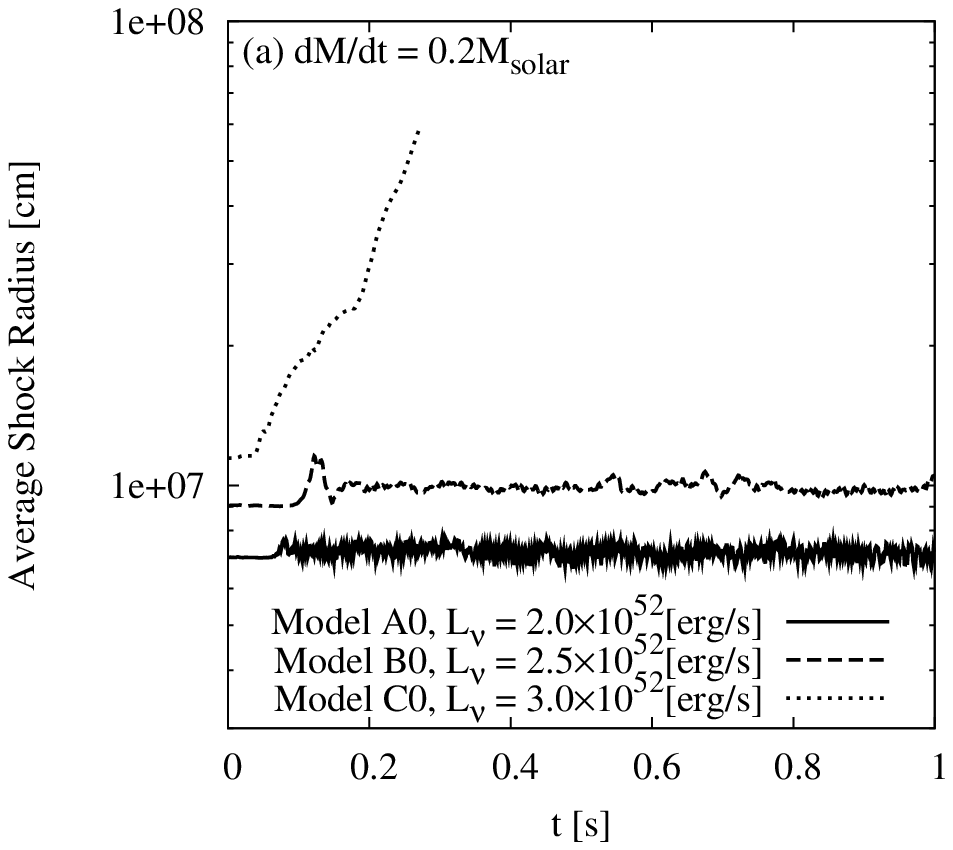}
\plotone{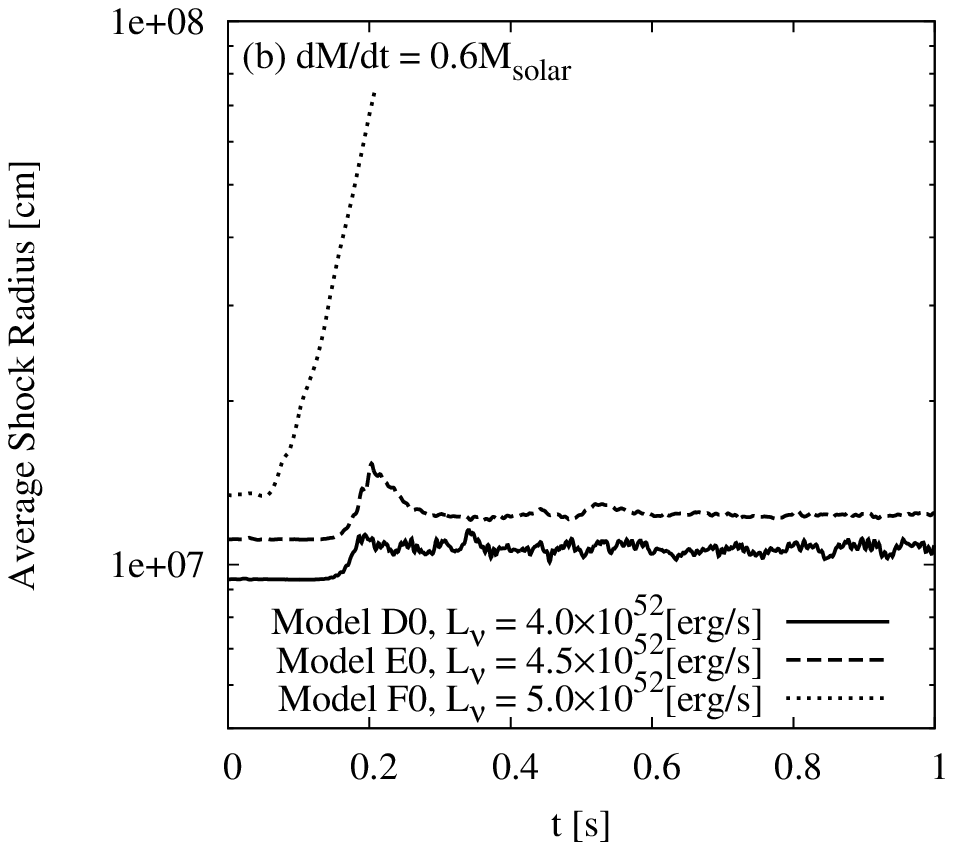}
\plotone{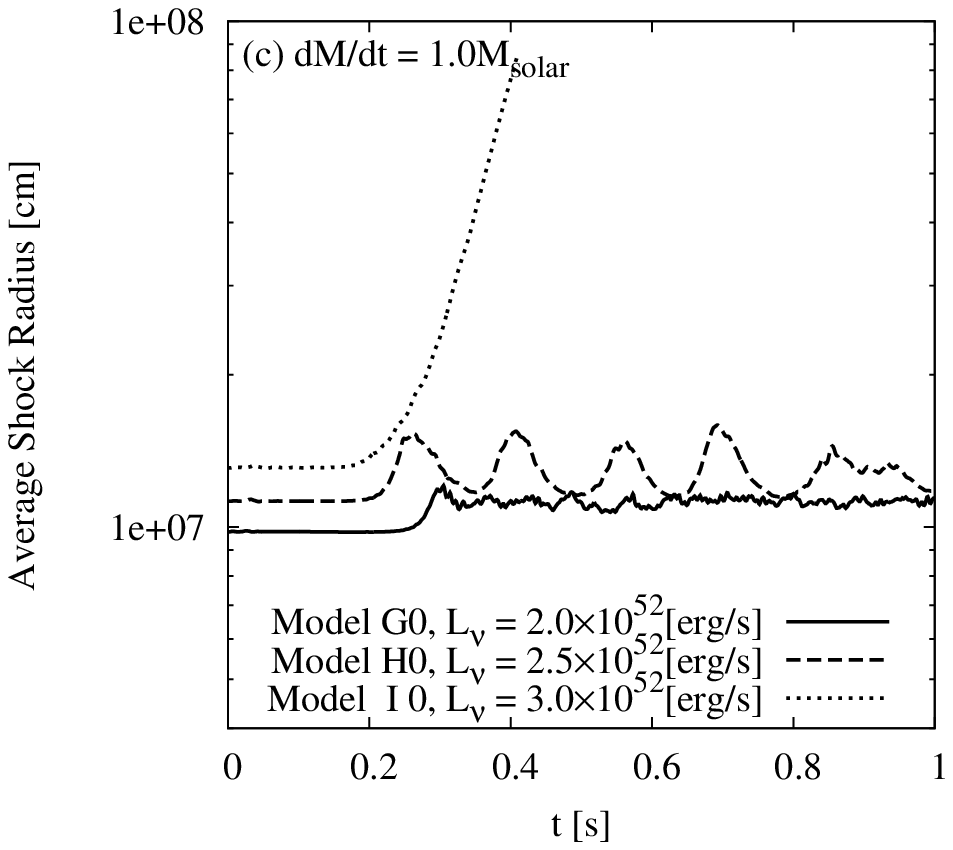}
\caption{
Time evolutions of the averaged shock radius for (a) $\dot{M} = 0.2 M_\odot$, (b) $0.6 M_\odot$, and (c) $1.0 M_\odot$. 
The solid, dashed, and dotted lines show the results for the low, middle, and high neutrino luminosities, respectively.
\label{fig_rad}}
\end{figure}

The time evolutions of the shock radius are shown in Figure~\ref{fig_rad}, obtained from the expansion coefficient $c_0^0$ in Eq.~(\ref{eq_c}).
The solid and dashed lines correspond to the non-explosion models, and the dotted lines to the explosion models.
In the non-explosion models, the average shock radius is larger for higher neutrino luminosities.
In the explosion models we observe the monotonic increase of the shock radius up to the outer boundary.

In both cases, the shock radius is almost constant for a certain period initially.
In this phase the perturbation grows exponentially from the initial small value with no substantial feed back to the background flow.
We refer to this period as the linear phase in this paper.
In the following period, which we call the semi-nonlinear phase, there is a discernible increase of the average shock radius that continues to the first peak in the non-explosion models.
As shown later in Figure~\ref{fig_mode}, the ending time of semi-nonlinear phase corresponds to the first turning point from exponential growth to its saturation or relaxation in the time evolution of mode amplitudes.
Nonlinear mode couplings cannot be ignored any more in this phase and a dominance of a particular mode becomes evident.
The semi-nonlinear phase is followed by the nonlinear phase, in which the hydrodynamical instabilities (convection and/or SASI) are saturated and quasi-steady states are established with some intermittent activities in the non-explosion models.
In the following we focus on the flow patterns realized in the semi-nonlinear and nonlinear phases.

\begin{figure}
\epsscale{.45}
\plotone{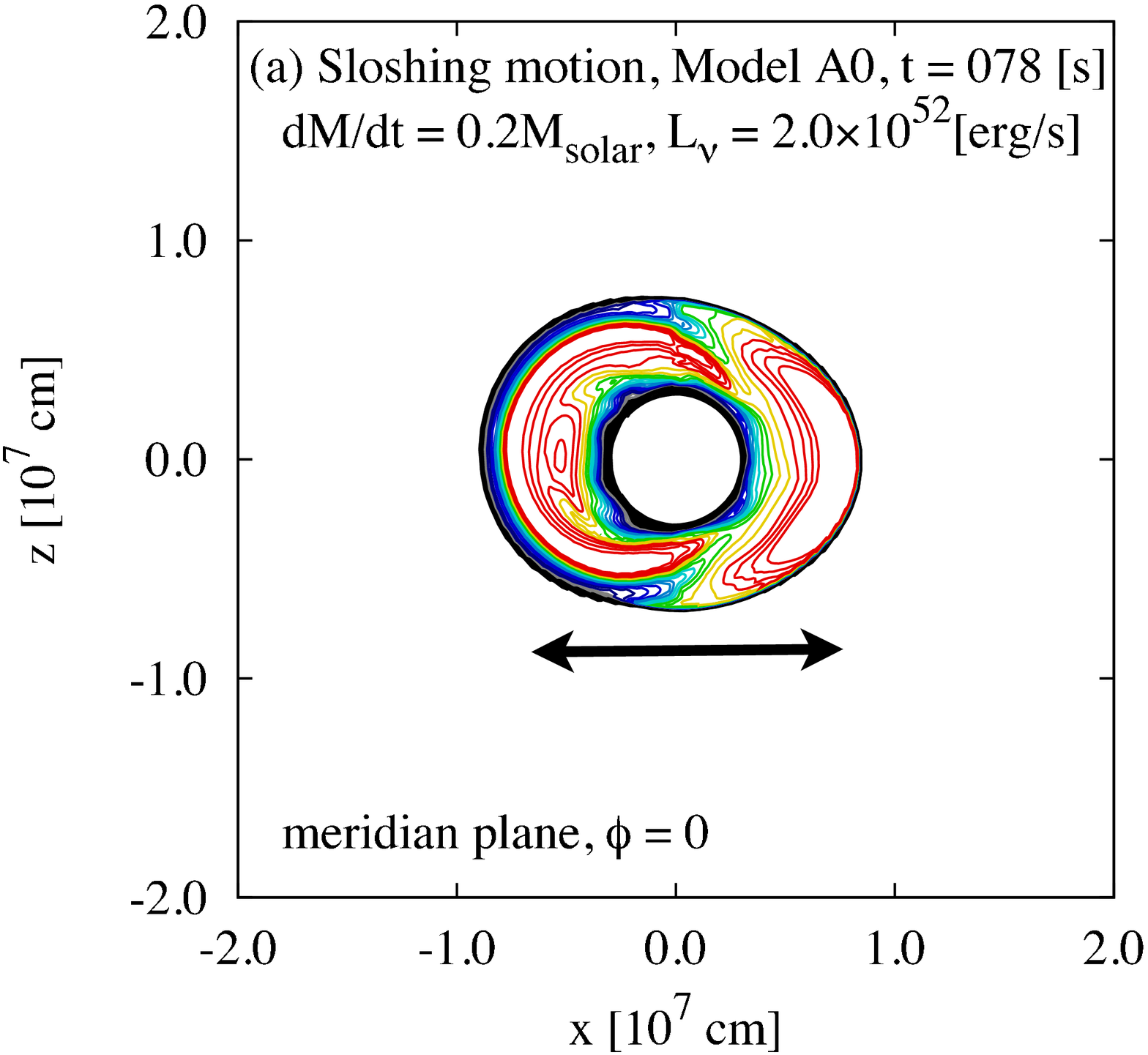}
\plotone{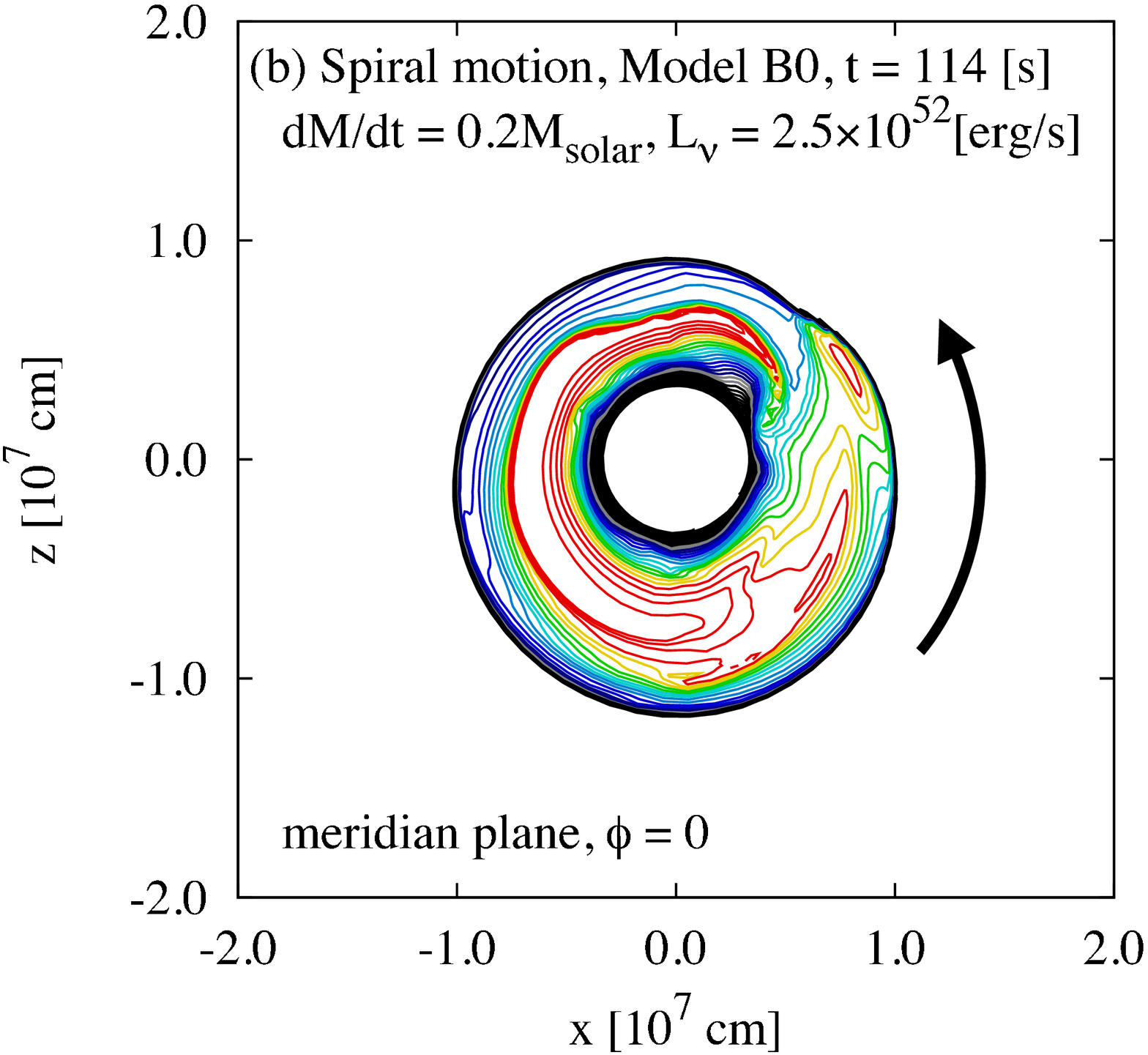}
\plotone{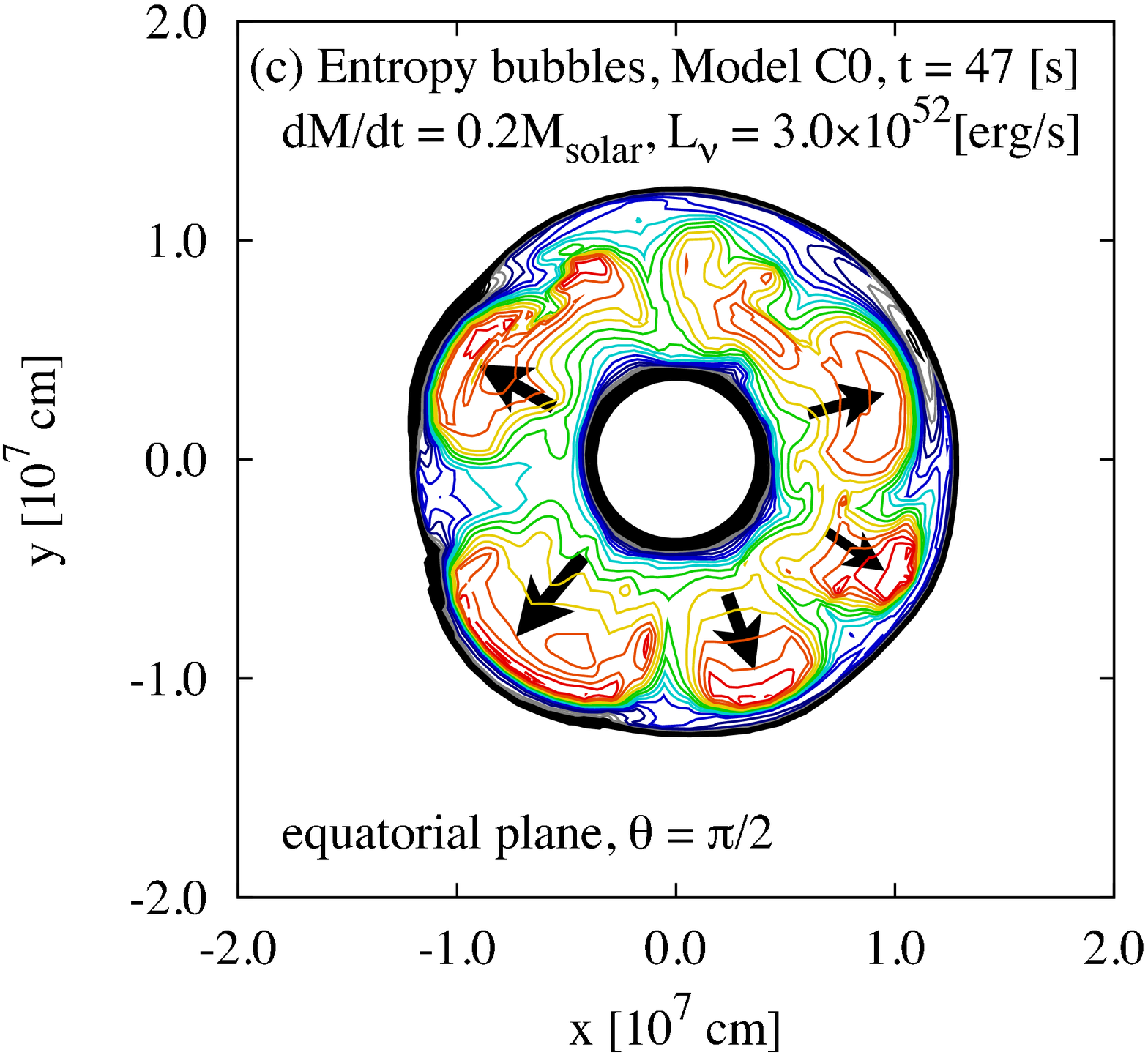}
\caption{
Snapshots of the entropy contour map in the meridian plane at $\phi=0$ for (a) sloshing motion, (b) spiral motion, and in the equatorial plane at $\theta=\pi/2$ for (c) entropy bubbles.
The entropy $S$ is in units of Boltzmann's constant $k_b$ per nucleon.
The contour levels are equally spaced in the range of $4 \le S \le 20$ with the increment of $\Delta S = 0.4$.
The contour lines of higher values are drawn in reddish colors, and those of lower ones are done in bluish colors.
The inner- and outermost contour lines agree with the surfaces of the proto-neutron star and the shock wave, respectively.
The arrows mean (a) the global oscillation of the shock wave in the indicated direction, (b) the global rotation of the shock wave deformations, and (c) the rising motions of buoyant bubbles toward the shock wave.
\label{fig_ent}}
\end{figure}

Here we introduce three well-known post-shock flow patterns: sloshing motion, spiral motion, and formation of multiple buoyant bubbles with high entropies.
Some snapshots of entropy distribution in a cross-section are presented as contours in Figure~\ref{fig_ent}.
The inner- and outermost contour lines correspond to the PNS surface and the shock front, respectively. 
The sloshing and spiral motions are accompanied by global deformations of shock front whereas the formation of buoyant bubbles generates shock deformations on a smaller scale.
The sloshing motion oscillates the shock front along a certain axis.
Figure~\ref{fig_ent}a shows an example of the sloshing motion, which occurs along the $x$ axis in this case.
The spiral motion is intrinsically non-axisymmetric with rotations of matter around a certain axis although the accretion is spherically symmetric outside the shock wave.
Figure~\ref{fig_ent}b presents such an example with the rotation axis almost perpendicular to the $x-z$ plane.
As shown in Figure~\ref{fig_ent}c on the other hand, the formation of high-entropy bubbles in the gain region tends to deform the shock wave rather locally.

\begin{figure}
\epsscale{.70}
\plotone{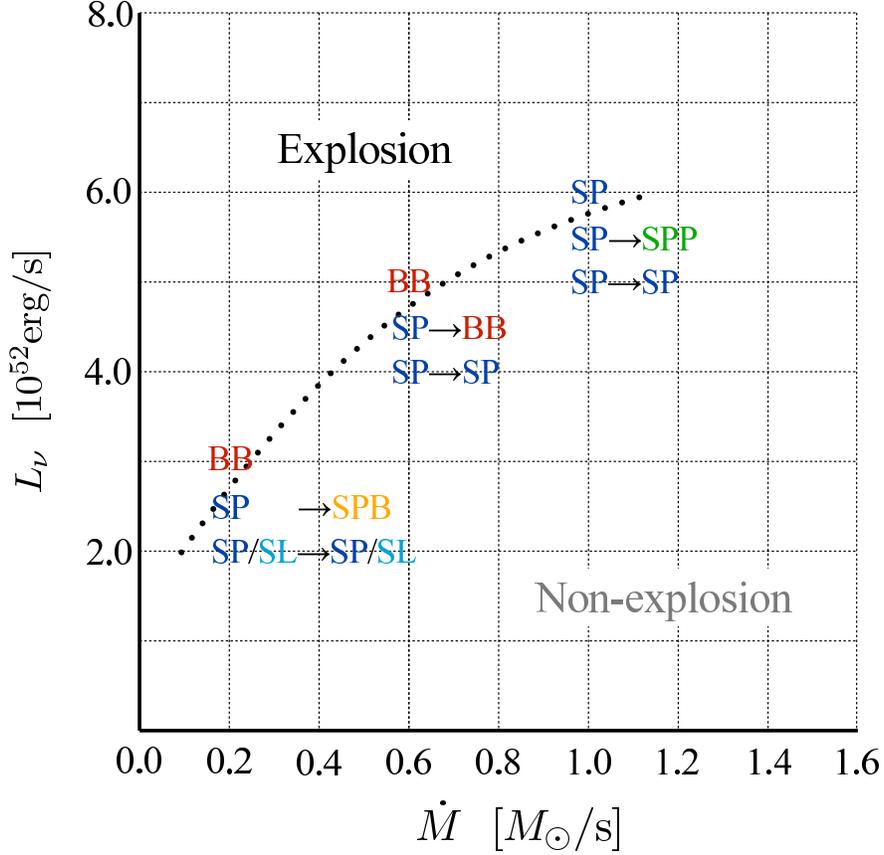}
\caption{
Phase diagram of flow patterns.
The critical line for explosion shown as a dotted curve is a rough guide drawn by hand.
Flow patterns are classified into sloshing motion (SL, light blue), spiral motion (SP, blue), formation of buoyant bubbles (BB, red), spiral motion with rising buoyant bubbles (SPB, yellow), and spiral motion with pulsating change of rotational velocities (SPP, green). 
See the body for more details. 
On the left of each arrow is given the flow pattern that appears in the semi-nonlinear phase, whereas the one in the nonlinear phase is indicated on the right.
For the explosion models, only the flow patterns in the semi-nonlinear phase are shown.
For the lowest accretion rate and neutrino luminosity, either SP or SL motion occurs, depending on the initial perturbation.
}
\label{fig_diag}
\end{figure}

We find from our simulations that the post-shock flows in the semi-nonlinear and nonlinear phases are classified either into the above three patterns: sloshing motion (abbreviated as SL in the following), spiral motion (SP) and buoyant-bubble formation (BB),
or into the following 
intermediate
patterns: spiral motion with bubble formation (SPB) and spiral motion with pulsating changes of rotational velocity (SPP).
Interestingly, the flow pattern is not always identical between the semi-nonlinear and nonlinear phases.
We hence summarize the results for the two phases separately in Table~\ref{tbl_kai}.
In order to facilitate the grasp of the trend in the pattern realization, we also show the same results in the $L_\nu-\dot{M}$ diagram in Figure~\ref{fig_diag}.
In this figure the flow patterns given on the left and right sides of arrows correspond to the flow patterns realized in the semi-nonlinear and nonlinear phases, respectively.
As discussed later, the pattern realized in the post-shock flow is rather robust, there are some exceptions, in which pattern change is observed when the initial perturbation is changed. 
In fact, two patterns put on the same site in Figure~\ref{fig_diag} implies that one of them is realized depending on the initial perturbation.
Note that the flow patterns are determined only in the semi-nonlinear phase for the explosion models, since the quasi-steady states are not established after the semi-nonlinear phase.

In Figure~\ref{fig_diag}, the dotted curve roughly indicates the critical luminosity estimated from the simulations in this study.
Although it agrees well with the results obtained by \citet{yamasaki06} for almost the same settings, it is generically higher than those found in more realistic simulations \citep{nordhaus10, hanke12, couch13a}.
Considering the simplifications adopted in this papers, we think that this much difference is not surprising.

We find that SL and SP tend to appear in the higher $\dot{M}$ and lower $L_\nu$, wheres BB prevails in the lower $\dot{M}$ and higher $L_\nu$.
This is consistent with \citet{burrows12} and \citet{hanke13}.
\citet{burrows12} performed parametrized simulations for a 15$M_\odot$ progenitor model
that vigorous sloshing motions are not in evidence in 3D.
Note that $\dot{M}$ is rather low in their models.
\citet{hanke13} demonstrated by their own 3D simulations for 25 and 27 $M_\odot$ progenitors that higher mass accretion rates lead to violent spiral motions, which are quenched when the accretion rate drops drastically with the infall of Si/SiO interface.

Looking at Figure~\ref{fig_diag} more in detail, we recognize the followings.
First, the indeterminacy of flow pattern is observed only for Model A, in which both $L_\nu$ and $\dot{M}$ are small. 
This may be a numerical artifact, since we find only SP in two higher-resolution simulations, as will be shown later in the section \ref{sec_gcheck}.
Considering the robustness in other cases, however, we think it is more likely that the indeterminacy is real and is yet another form of 
intermediate
case.
Second, only near the critical neutrino luminosity, we find different flow patterns between the semi-nonlinear and nonlinear phases.
Although SP always prevails in the semi-nonlinear phase, in the nonlinear phase BB is mixed with SP to form SPB in Model B and is fully dominant in Model E whereas another 
intermediate
pattern SPP is realized in Model H.
The characteristics of each flow pattern and the diversity in the nonlinear phase near the critical neutrino luminosity are discussed in the next section in more detail.

\subsection{Dynamical Features of Flow Patterns \label{sec_dyn}}

In this section, we discuss the dynamical features, based on which we classify the flow patterns.
We firstly focus on the semi-nonlinear phase, and then turn to the nonlinear phase.

\subsubsection{Semi-Nonlinear Phase \label{sec_semi}}

In this paper, the semi-nonlinear phase is defined to be the period following the linear growth phase, in which there is a discernible increase of the average shock radius up to the first turning point from exponential growth to its saturation or relaxation in the time evolution of mode amplitudes.
In this phase, nonlinear mode-couplings are no longer ignored and a dominant mode emerges as a consequence, and we find that one of the three basic flow patterns,
sloshing motion (SL), spiral motion (SP), and high-entropy buoyant-bubble formation (BB), is dominant, depending on the mass accretion rate and the neutrino luminosity.
We pay attention to the features in the temporal evolution of each flow pattern in the following.

\begin{figure}
\epsscale{0.90}
\plotone{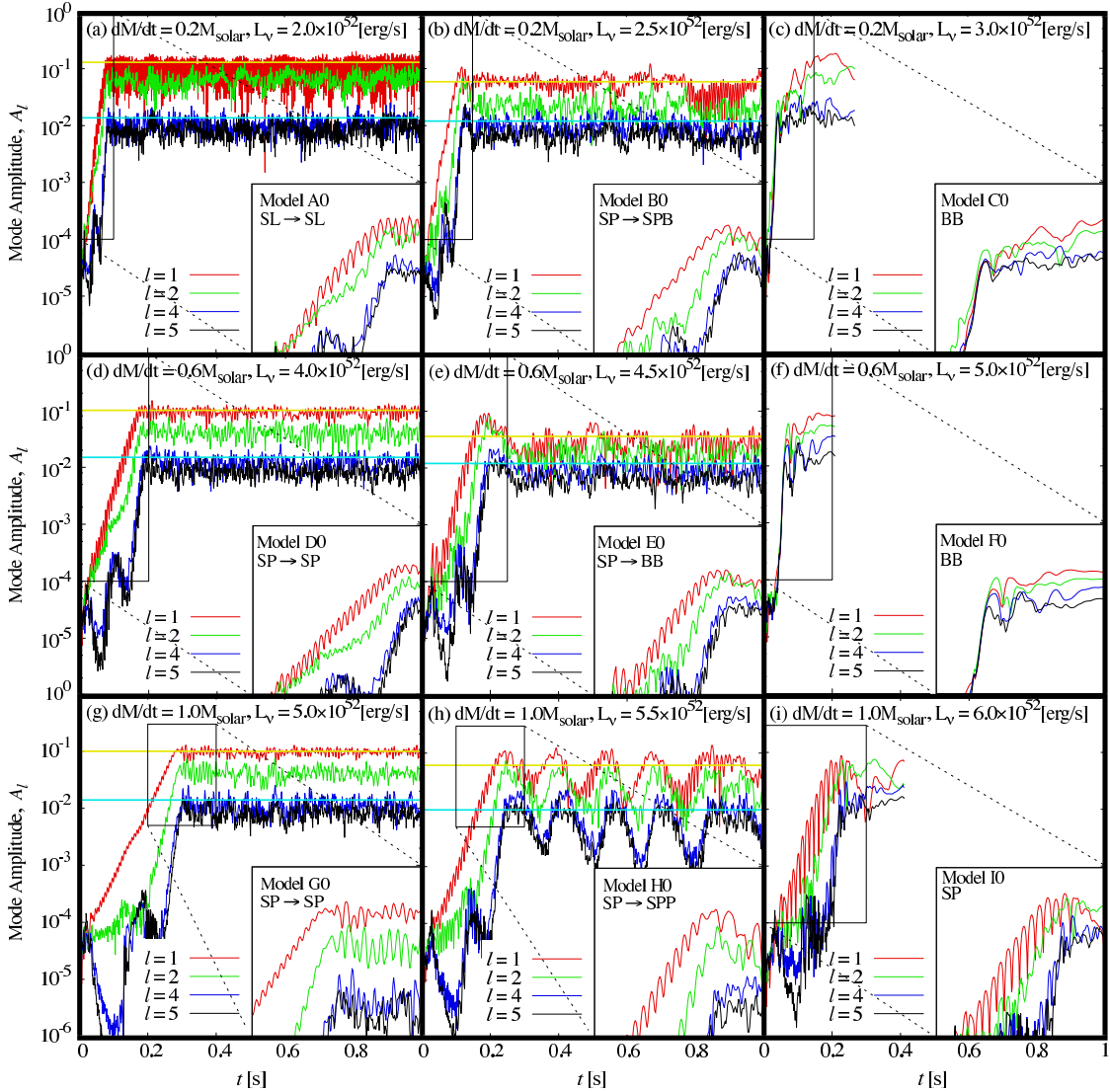}
\caption{
Time evolutions of the normalized mode amplitudes in the shock deformation.
The results for the same mass accretion rate are put in the same rows.
The neutrino luminosities go up from left to right.
The yellow lines correspond to the combined mode amplitudes of $l=1, 2$ averaged from 400ms to 1000ms, whereas the blue lines stand for those of $l=4, 5$.
The insets are the zoom-ups of the indicated portions.
\label{fig_mode}}
\end{figure}

Figure~\ref{fig_mode} shows the time evolutions of the mode amplitudes introduced in Section 2.2 as Eq.~(\ref{eq_A}).
The inset in each figure is a zoom-up of the semi-nonlinear phase.
We clearly see the dominance of the $l=1$ mode (SL or SP) with a periodic oscillation superimposed on an exponential growth in figures~\ref{fig_mode}a, \ref{fig_mode}b, \ref{fig_mode}d, \ref{fig_mode}e, \ref{fig_mode}g, \ref{fig_mode}h and \ref{fig_mode}i.
In figures~\ref{fig_mode}c and \ref{fig_mode}f, on the other hand, several modes grow monotonically with similar growth rates, which is reminiscent of the non-oscillatory unstable modes
that are identified as convective modes in their linear analysis by \citet{yamasaki07}.
We hence identify the latter two cases as BB, which is confirmed by the entropy distributions in the post-shock flows.

In the literature the $\chi$ parameter proposed by \citet{foglizzo06} is frequently used as a measure of convective instability.
Since this criterion is based on the linear analysis, it is rather ambiguous for what flow the $\chi$ parameter should be calculated except for the linear phase.
For the analysis of the semi-nonlinear phase, we employ the unperturbed states.
The values of $\chi$, calculated by Eqs.~(\ref{eq_chi}) and (\ref{eq_n}), are listed in the Table \ref{tbl_kai}.
It is found that BB is realized for $\chi \gtrsim 4$ and SP or SL is observed otherwise, which seems to be consistent with the original criterion $\chi > 3$ although the critical value may be a little larger in our models.
It is also noted that the critical $\chi$ values in our models are much larger than that found in \citet{yamasaki07}, the reason of which is not clear for the moment.

SL and SP may be discriminated by the temporal evolution of the orientation of the instantaneous rotation axis for the post-shock flow.
Note that in the linear phase only the pattern is rotating and matter has a negligible angular momentum in the spiral SASI.
In the semi-nonlinear phase (and non-linear phase as well), on the other hands, matter also rotates in the same direction as the pattern \citep{blondin07}.
The sloshing motion does not generate angular momentum in principle.
We hence expect that the instantaneous rotation axis is stable in SP with the orientation being fixed longer than the instantaneous rotation period whereas it changes orientation stochastically in SL.
Such a dichotomy is actually observed in Figure \ref{fig_direc}, in which we show the orientation of the instantaneous rotation axis, which is denoted by $\theta$ and $\phi$ in the polar coordinate system and is calculated from the instantaneous angular momentum integrated over the post-shock flow.
In panels~\ref{fig_direc}b, \ref{fig_direc}d, \ref{fig_direc}e, \ref{fig_direc}g, \ref{fig_direc}h and \ref{fig_direc}i, we see that the orientation of the instantaneous rotation axis remains almost fixed for a while and changes its directions from time to time rather suddenly, which is the feature we expect for SP. 
In panel~\ref{fig_direc}a, on the other hand, we observe a random walk of the rotation axis, which we suppose to characterize SL. 
We hence identify the flow patterns for B0, D0, E0, G0, H0 and I0 as SP and that for A0 as SL in the semi-nonlinear phase.

\subsubsection{Nonlinear Phase \label{sec_nonlin}}

After the instability grows up in the linear and semi-nonlinear phases, it enters the fully nonlinear phase and is saturated.
It is a non-trivial task to classify various flow patterns in this phase objectively, but the ratio of the time-averaged mode amplitude of $l=1, 2$ to that of $l=4, 5$, may be useful, since strong sloshing and spiral motions are expected to produce large saturation amplitudes of low $l$ modes, whereas multiple high-entropy buoyant-bubble formations will generate high $l$ modes.
Moreover, we find mixed patterns (SPP and SPB) in the nonlinear phase, which have intermediate ratios as expected.
Details will follow shortly.

\begin{figure}
\epsscale{0.90}
\plotone{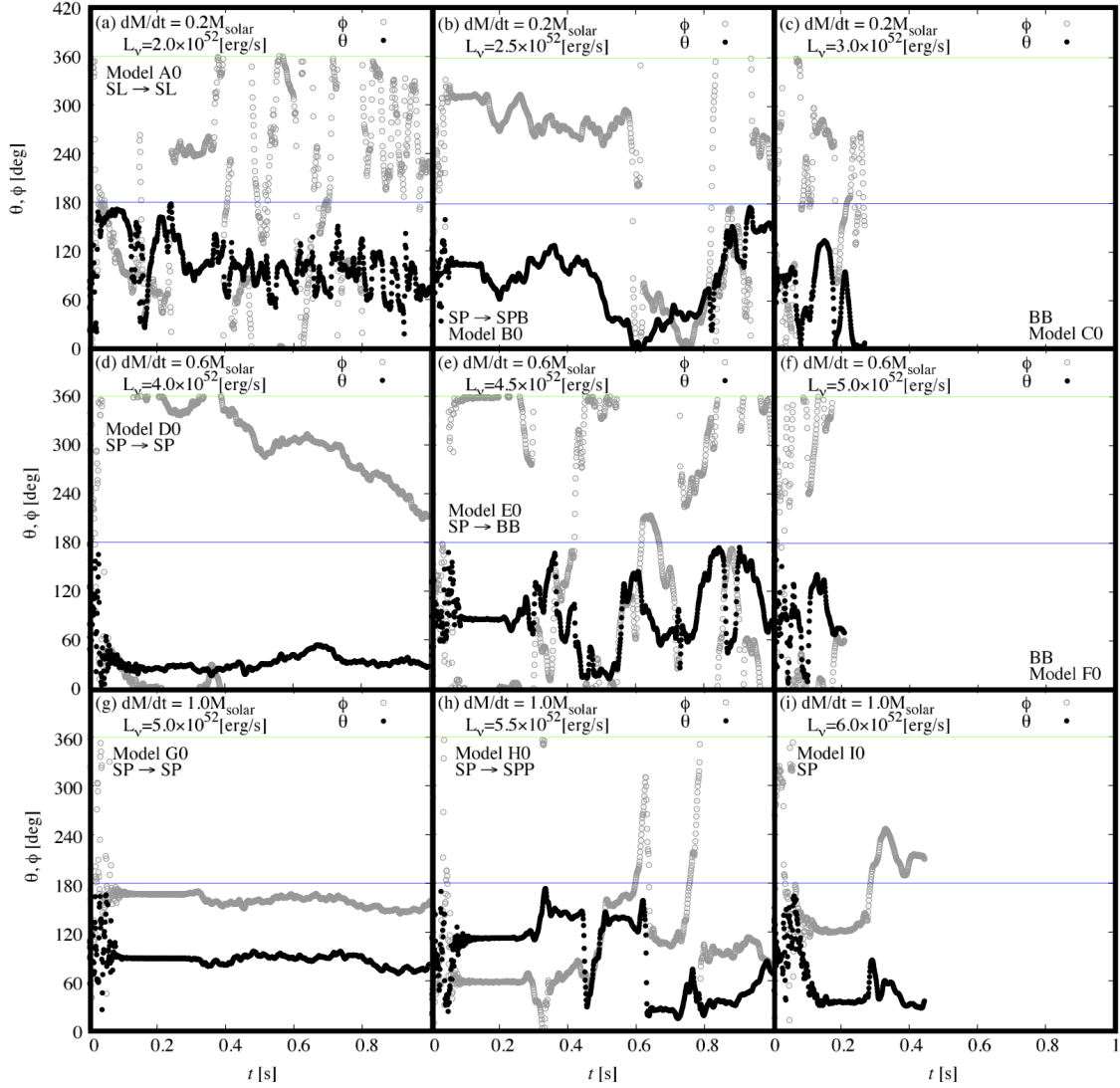}
\caption{
Time evolutions of the orientation of the rotation axis determined from the instantaneous angular momentum integrated over the post-shock flow.
The angles $\theta$ (black filled-circle) and $\phi$ (gray open-circle) indicating the orientation, are the spherical coordinates.
The blue and green lines correspond to $\pi$ and 2$\pi$, respectively.
\label{fig_direc}}
\end{figure}

\begin{figure}
\epsscale{0.90}
\plotone{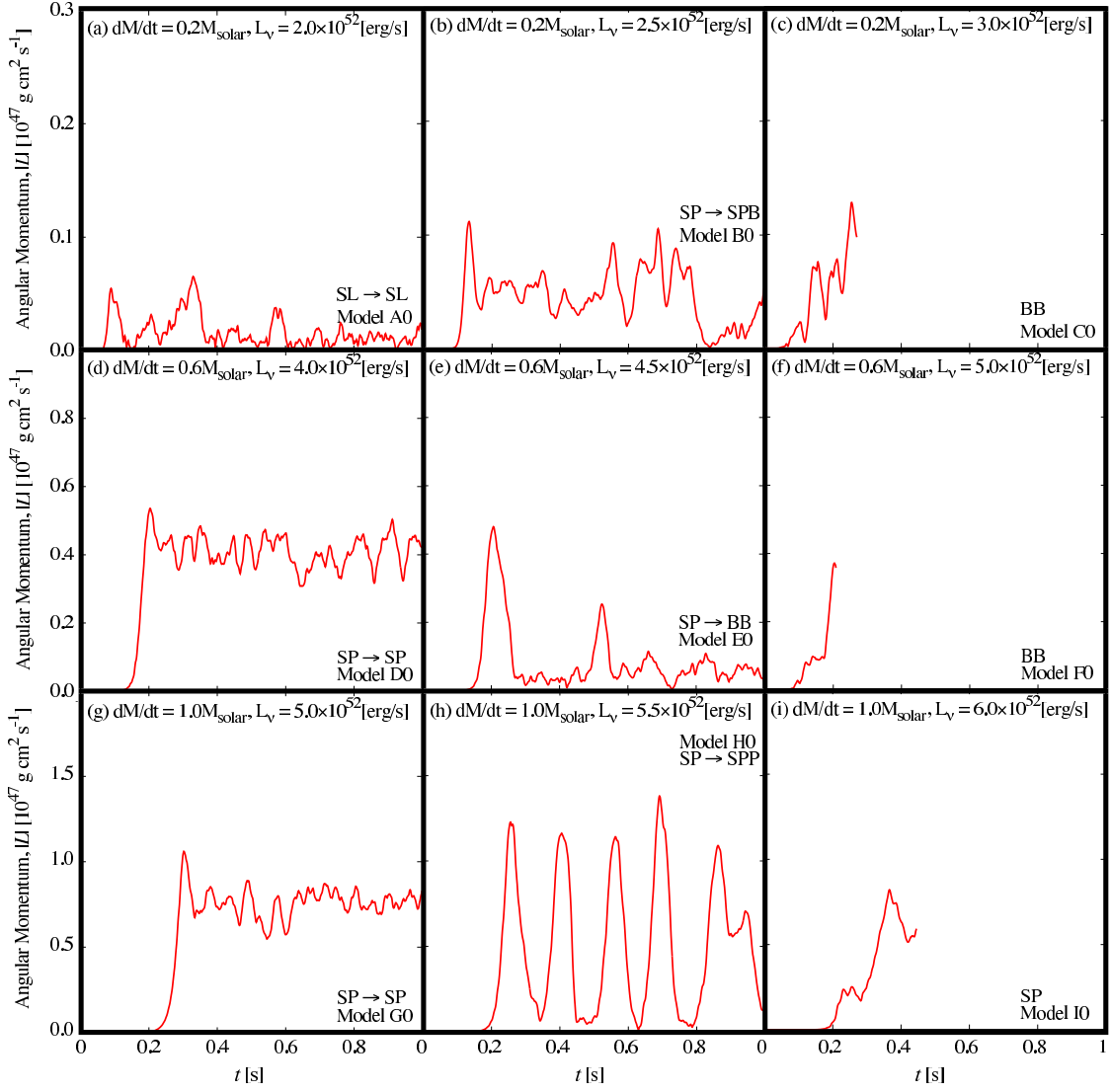}
\caption{
Time evolutions of the magnitude of angular momentum integrated over the post-shock flows.
\label{fig_angmom}}
\end{figure}

First, we discuss the sloshing pattern (SL).
High saturation amplitudes of low $l$ modes (Fig.~\ref{fig_mode}a), extremely low angular momenta (Fig.~\ref{fig_angmom}a), and unstable rotation axis (Fig.~\ref{fig_direc}a) indicate that the sloshing motion is dominant for model A0.
In fact, the ratio of $\bar{A}_{1,2}$ to $\bar{A}_{4,5}$ is pretty large, $\sim 10$, as shown in Figure~\ref{fig_mode}, in which we present the temporal evolutions of various modes as well.
Figure~\ref{fig_angmom} shows the time evolutions of the magnitude of the angular momentum integrated over the post-shock flow.
The direction of sloshing motion changes from time to time in 3D as observed around $t=300$ms in Model A0.
It is important that SL is maintained after such rather violent transitions.

Next, we focus on the spiral pattern (SP).
High saturation amplitudes of low $l$ modes (Fig.~\ref{fig_mode}d and \ref{fig_mode}g) are common with SL, but high angular momenta (Fig.~\ref{fig_angmom}d and \ref{fig_angmom}g) as well as stable rotation axis (Fig.~\ref{fig_direc}d and \ref{fig_direc}g) characterize SP.
The rotation axis shifts gradually for model D0 and it is almost fixed during $\sim$1s for model G0.
The magnitude of anugular momentum in the post-shock flow is proportional to the mass accretion rate in Models D and G, for which the shock radii are almost identical.
The results are consistent with the analytical study by \citet{guilet13}.
Model H0 is a peculiar case.
The magnitude of angular momentum changes periodically, i.e. rapid and slow rotations occur alternatively with a period of $\sim$150ms (Fig.~\ref{fig_angmom}h).
The orientation of rotation axis varies stochastically at the local minimum of the angular momentum (Fig.~\ref{fig_direc}h and \ref{fig_angmom}h).
We hence refer to this flow pattern as the spiral motions with pulsating change of rotational velocities (SPP).
Reflecting this situation, the amplitudes of all modes including $l=0$ oscillate in phase (Fig.~\ref{fig_rad}c and Fig.~\ref{fig_mode}h).
Such large time-variations of $l=2$ mode may give additional features to the gravitational wave signal.
Interestingly, \cite{fernandez09b} reported a similar phenomenon in their axisymmetric 2D simulations of post-shock flows, which hence lack spiral motions.
In fact, they found intermittent sloshing motions, depending on the dissociation energy.
This might give us a clue to the understanding of SPP.

We now turn to BB, buoyant-bubble formations.
Model E0 is prototypical.
After the amplitude of the dominant spiral mode reaches $\sim$10\% of the shock radius, the flow pattern changes from SP to BB and the amplitudes of various modes are reduced to $\sim$4\% level (Fig.~\ref{fig_mode}e).
It is observed in this nonlinear phase that multiple buoyant-bubbles are formed repeatedly and collide with the shock wave randomly.
As a result, large-scale deformations of the shock wave as we see them in SL or SP are suppressed in BB.
This is most clearly indicated by the ratio of $\bar{A}_{1,2}$ to $\bar{A}_{4,5}$, which is much smaller than those for SL or SP.
Moreover, buoyant bubbles disrupt the regular circulations around PNS.
As a result, the angular momentum is not so high (Fig.~\ref{fig_angmom}e), and the orientation of rotation axis changes rapidly in a short period (Fig.~\ref{fig_direc}e).
This unstable behavior of the rotation axis is another characteristic that we adopt in judging the flow pattern as BB.
Our previous 3D simulations did not generate strong spiral motions as found by \citet{blondin07} and the saturation amplitudes, a few percents, were much smaller than theirs.
The reason for this apparent discrepancy is now clear: we worked in the BB regime.

\begin{figure}
\epsscale{0.70}
\plotone{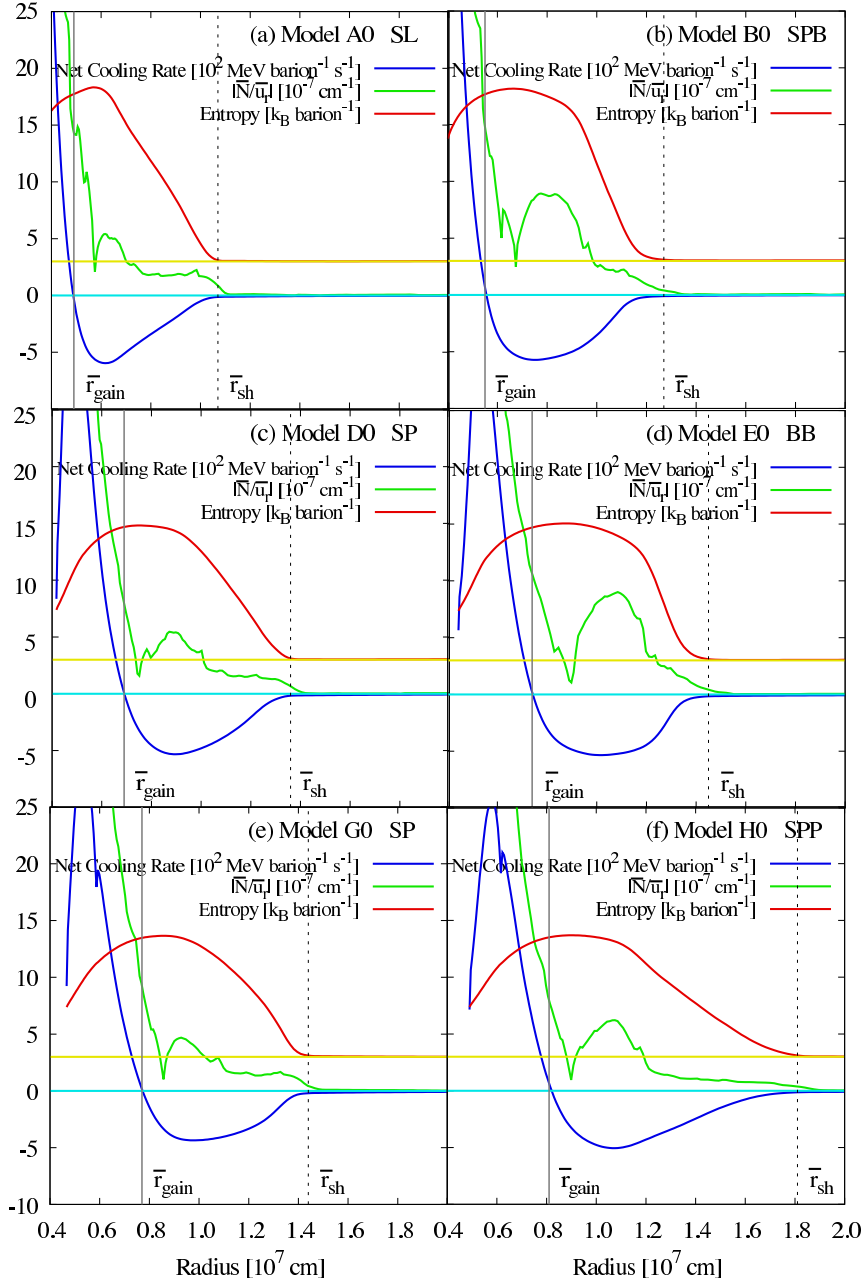}
\caption{
The radial distributions of entropy,
net cooling rate,
and $|\bar{N}/\bar{u}_r|$
in the angle-averaged flows in the quasi-steady nonlinear phase for non-explosion models.
The yellow lines indicate the position of 3, the value of entropy in the upstream flow whereas the light blue lines show the position of 0.
The average gain and shock radii are denoted by $\bar{r}_{\rm gain}$ and $\bar{r}_{\rm sh}$, respectively.
\label{fig_aver}}
\end{figure}

In applying the criterion for convection with the $\chi$ parameter to the nonlinear phase of non-explosion models, we employ the time- and angle-averaged flows that are evaluated according to Eqs.~(\ref{eq_tb}) and (\ref{eq_ab}) at $t_s=0.4$s and $t_e=1.0$s.
In Figure~\ref{fig_aver}, we show the entropy $\bar{S}$, the net cooling rate $\bar{Q}_\nu$, and the ratio of the Brunt-V\"{a}is\"{a}l\"{a} frequency to the radial velocity $|\bar{N}/\bar{u}_r|$ as a function of radius for Models A0, B0, D0, E0, G0, and I0 in Figure~\ref{fig_aver}.
The average gain radius $\bar{r}_{\rm gain}$ is located at $\bar{Q}_\nu=0$, and the 
maximum
shock radius $\bar{r}_{\rm sh}$ is defined to be the radius with $\bar{S}= 3.1$.
One observes that a larger portion of the gain region has rather flat entropy distributions in BB and SPB than in other flow patterns.
This region corresponds to convective activities.
One also finds that the ratio of the Brunt-V\"{a}is\"{a}l\"{a} frequency to the radial velocity becomes large in this region.
As a consequence, $\bar{\chi}$ obtained by the integration of this ratio over the gain region is greater for BB than for SP/SL.
The numerical values are listed in Table~\ref{tbl_kai}.
It is recognized that $\bar{\chi}$ for SPB is a bit larger than that for BB, which is at odds with the expectation.
This is likely due to insufficient grid resolutions, though.
As a matter of fact, $\bar{\chi}$ for BB tends to increase with the spatial resolution as shown in Table~\ref{tbl_kai_check} and will be discussed later in section \ref{sec_gcheck} more in detail.

Next, we consider the 
intermediate
pattern between SP and BB, that is, SPB.
Model B0 is the prototype.
The ratio of $\bar{A}_{1, 2}$ to $\bar{A}_{4, 5}$ for this model lies between the values for SP and BB as expected (Fig.~\ref{fig_mode}b).
When the angular momentum is relatively large (Fig.~\ref{fig_angmom}b), the rotation axis gradually changes its orientation as in SP for most of time (Fig.~\ref{fig_direc}b).
When the angular momentum becomes small as observed at $t\sim0.6$s and $t\sim0.8-1.0$s (Fig.~\ref{fig_angmom}b), however, the rotation axis is turned as in BB (Fig.~\ref{fig_direc}b).
In these periods, spiral features disappear and convective ones emerge instead.
We hence classify this flow pattern in model B0 as SPB.

Finally, we discuss the explosion models: C0, F0 and I0.
In these models it takes the shock wave only a few hundred milliseconds to reach the outer boundary of the computational domain, which is located at $r\sim 500-1000$km.
We may extract some more information from the late evolutions of these models, though.
The low $l$ modes tend to have greater amplitudes and longer oscillation periods in the explosion models than in the non-explosion models (Fig.~\ref{fig_mode}c, \ref{fig_mode}f and \ref{fig_mode}i).
This feature was reported in other works \citep[e.g.,][]{burrows12}.
The angular momentum generated in the spiral modes tend to be larger as the shock radius increases in the explosion models.
It is also positively correlated with the mass accretion rate (Fig.~\ref{fig_angmom}c, \ref{fig_angmom}f and \ref{fig_angmom}i).
These are consistent with the results in \citet{guilet13}.
The amplification of angular momentum in the nonlinear phase was also found in \citet{blondin07}.
The unstable behavior of the rotation axis is observed in all the explosion models, it is slower in Model I0 than in Models C0 and F0 (Fig.~\ref{fig_direc}c, \ref{fig_direc}f and \ref{fig_direc}i).

\subsection{Different Realizations \label{sec_rcheck}}

\begin{figure}
\epsscale{0.90}
\plotone{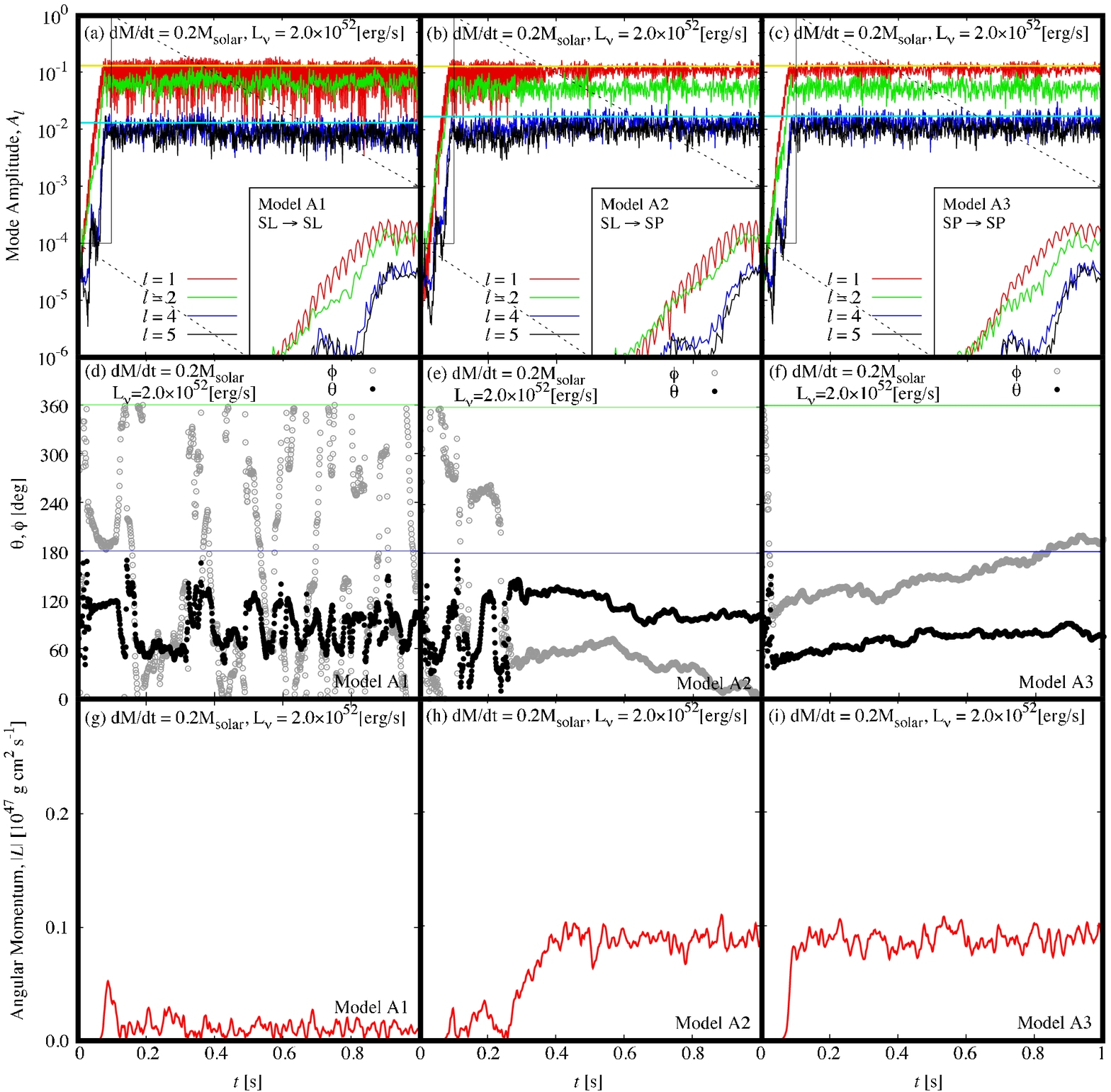}
\caption{
Time evolutions of the normalized mode amplitudes, the orientations of the rotation axis, and the magnitudes of angular momentum for Models A1, A2, and A3.
\label{fig_modelA}}
\end{figure}

\begin{figure}
\epsscale{0.90}
\plotone{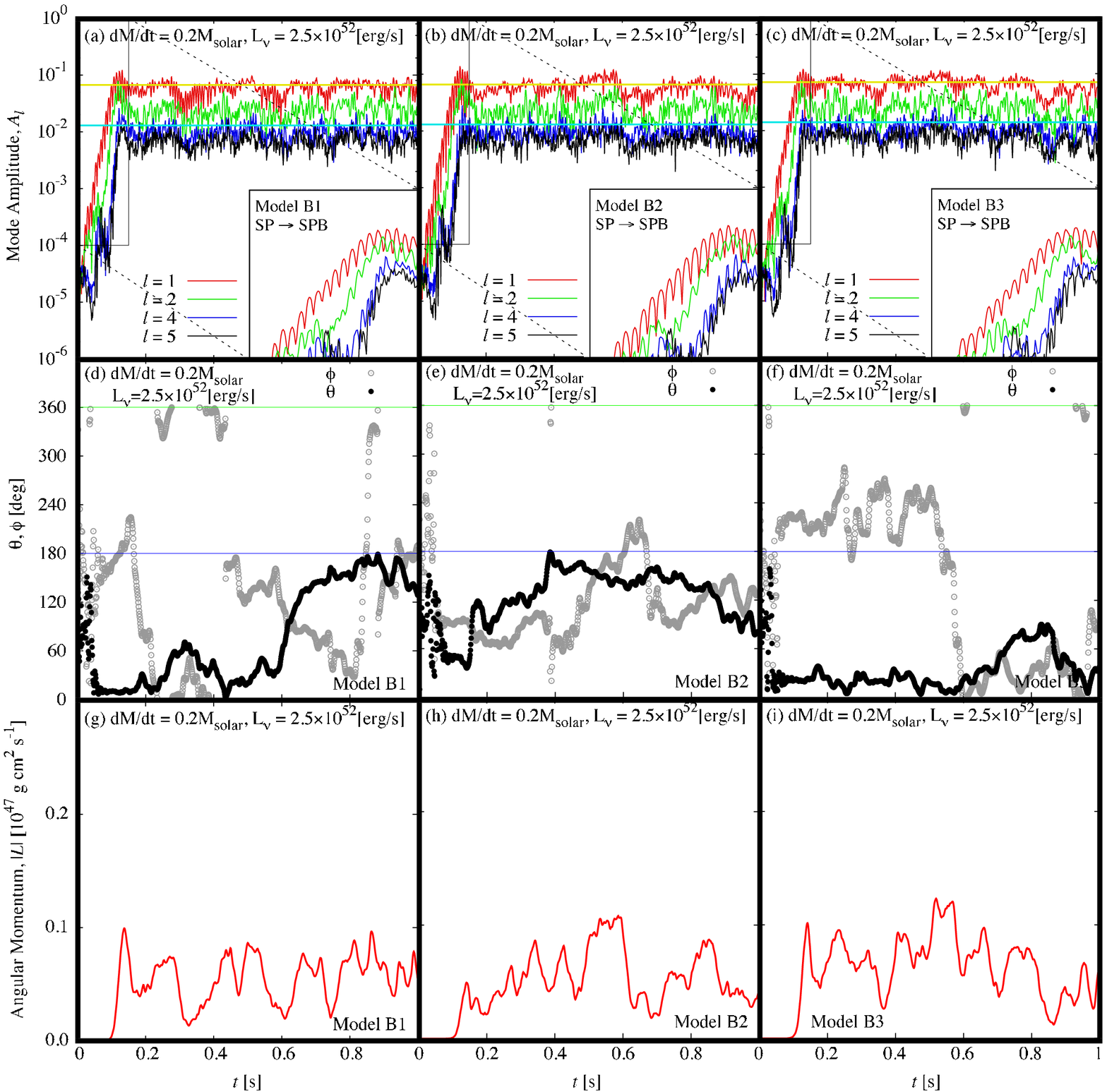}
\caption{
Time evolutions of the normalized mode amplitudes, the orientations of the rotation axis, and the magnitudes of angular momentum for Models B1, B2, and B3.
\label{fig_modelB}}
\end{figure}

\begin{figure}
\epsscale{0.90}
\plotone{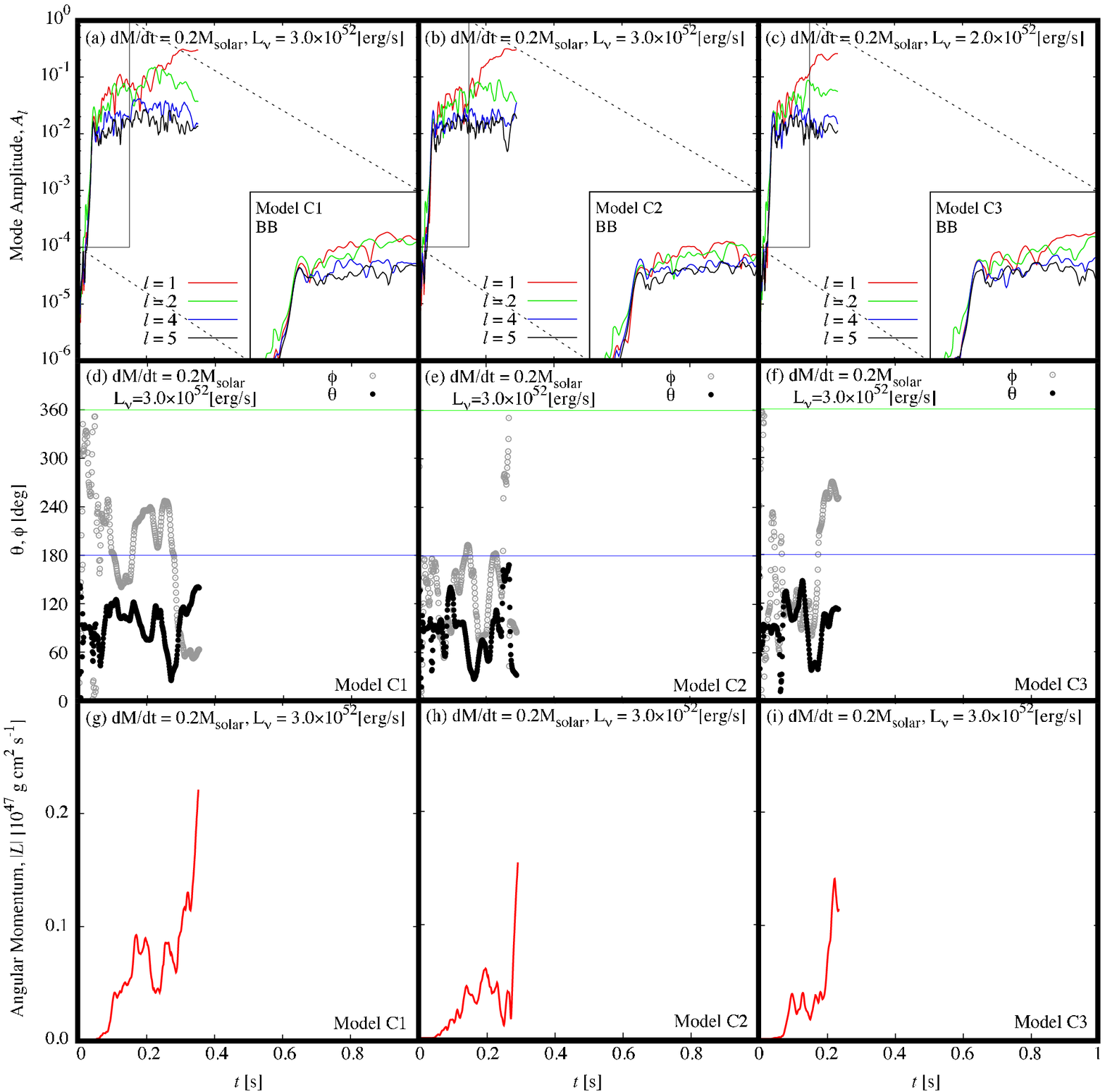}
\caption{
Time evolutions of the normalized mode amplitudes, the orientations of the rotation axis, and the magnitudes of angular momentum for Models C1, C2, and C3.
\label{fig_modelC}}
\end{figure}

\begin{figure}
\epsscale{0.90}
\plotone{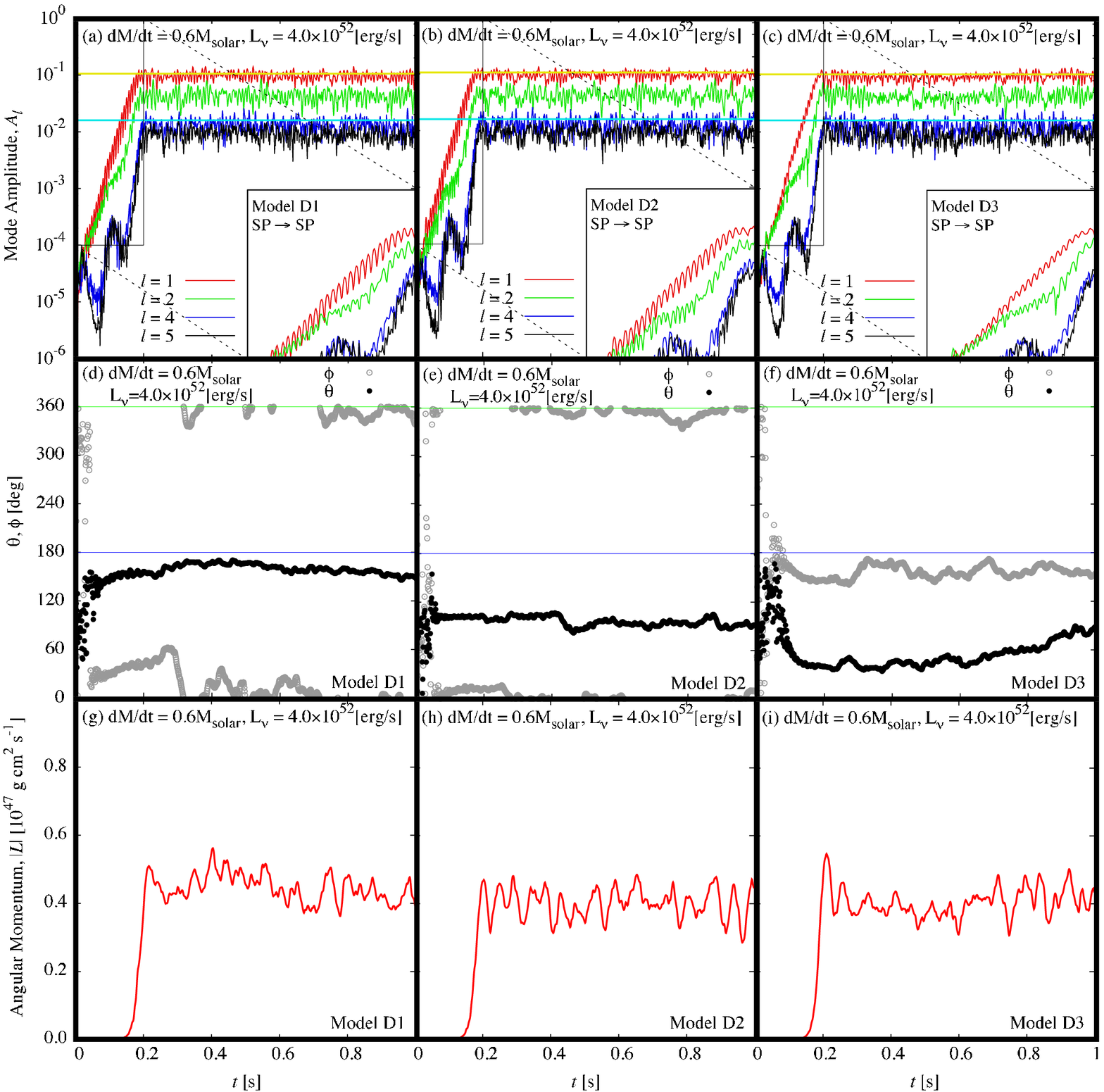}
\caption{
Time evolutions of the normalized mode amplitudes, the orientations of the rotation axis, and the magnitudes of angular momentum for Models D1, D2, and D3.
\label{fig_modelD}}
\end{figure}

\begin{figure}
\epsscale{0.90}
\plotone{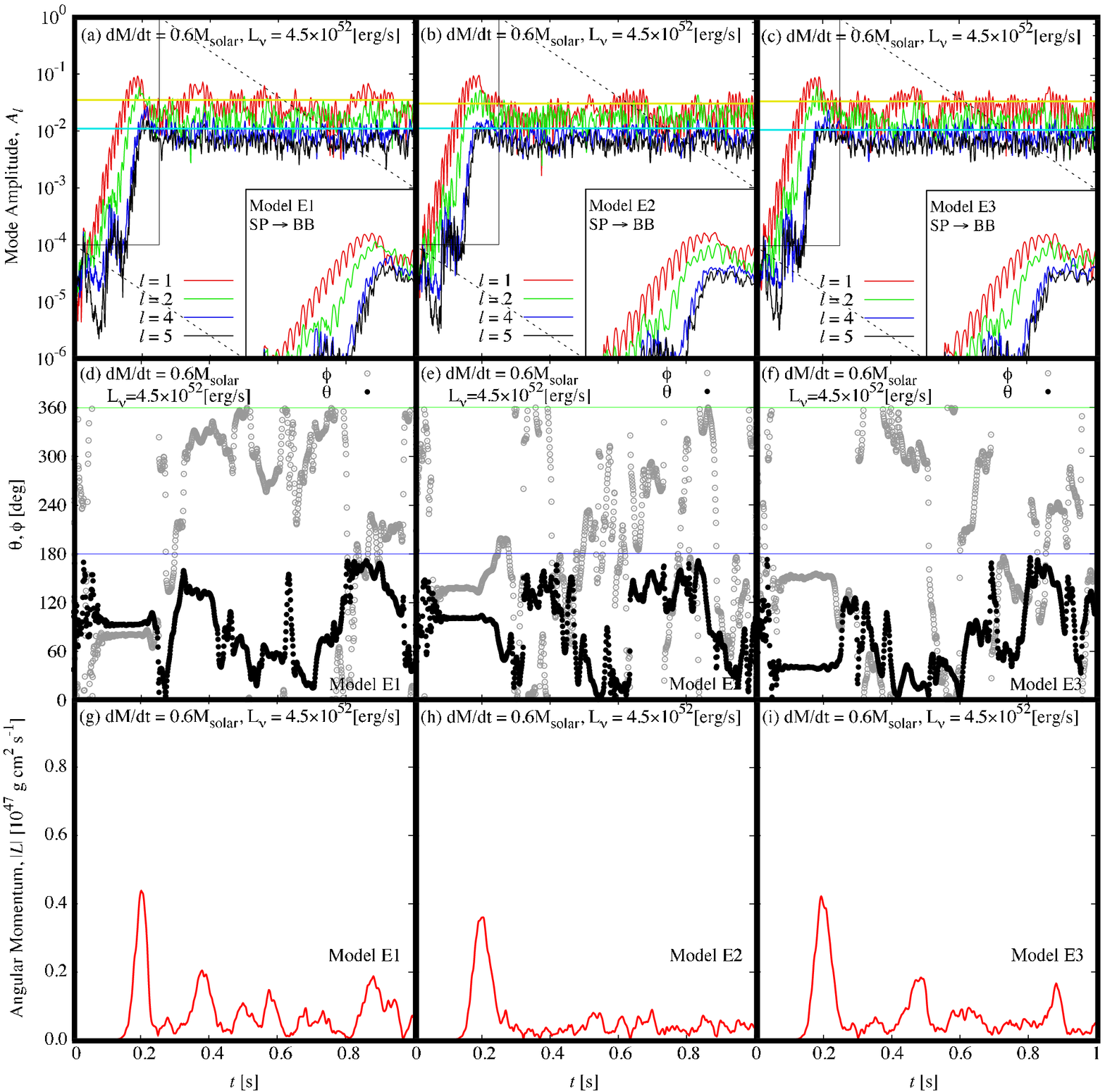}
\caption{
Time evolutions of the normalized mode amplitudes, the orientations of the rotation axis, and the magnitudes of angular momentum for Models E1, E2, and E3.
\label{fig_modelE}}
\end{figure}

\begin{figure}
\epsscale{0.90}
\plotone{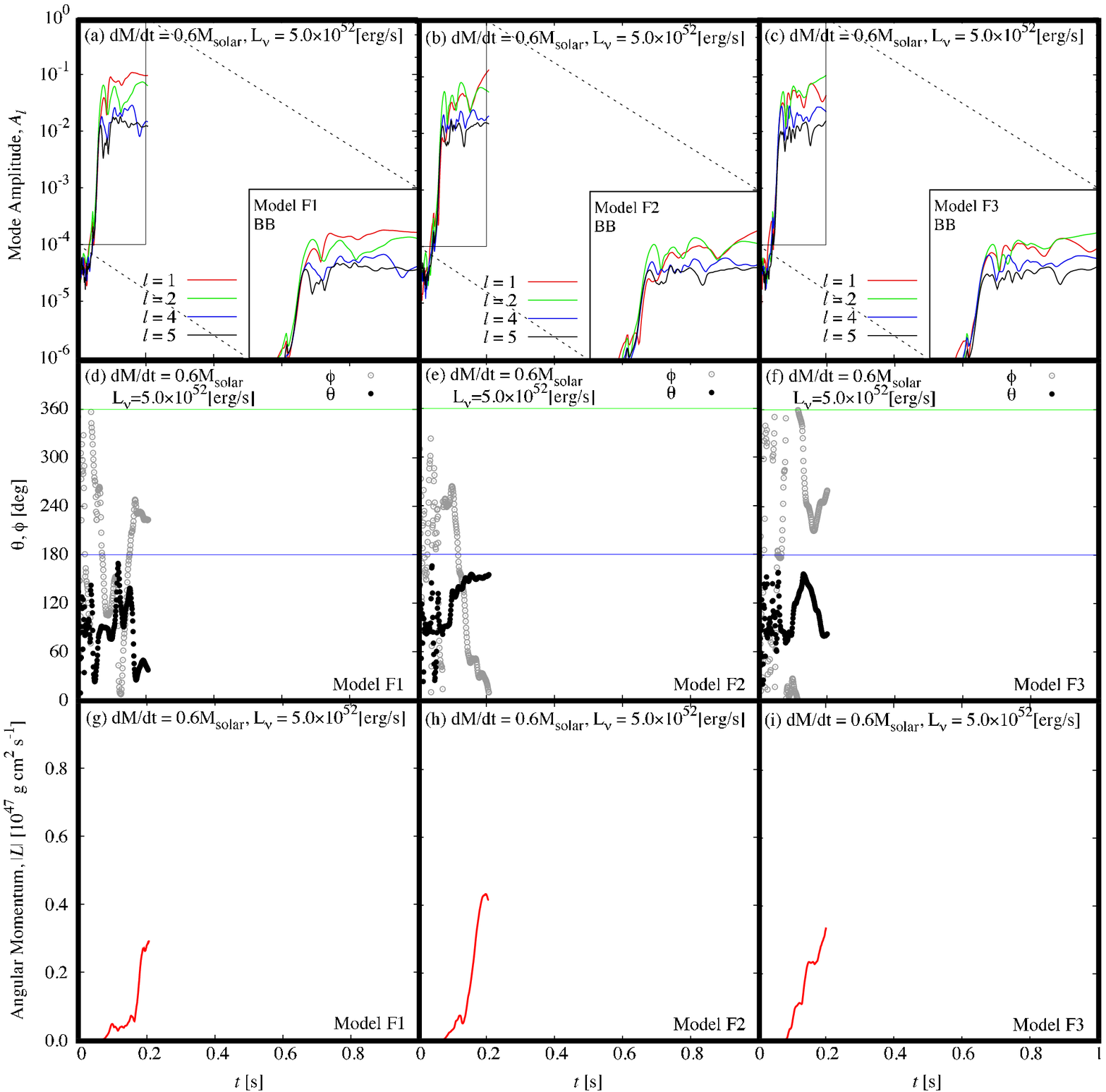}
\caption{
Time evolutions of the normalized mode amplitudes, the orientations of the rotation axis, and the magnitudes of angular momentum for Models F1, F2, and F3.
\label{fig_modelF}}
\end{figure}

\begin{figure}
\epsscale{0.90}
\plotone{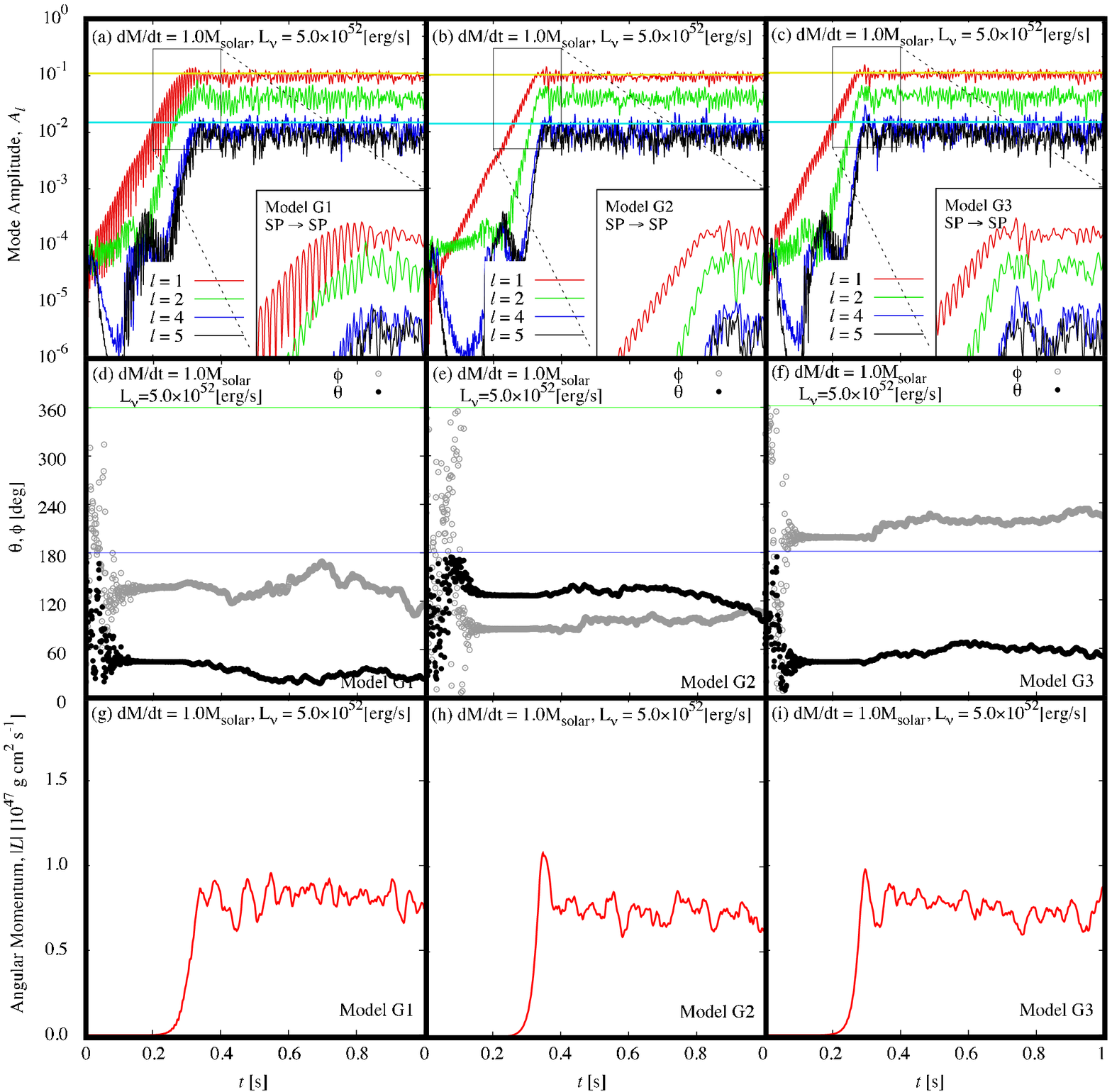}
\caption{
Time evolutions of the normalized mode amplitudes, the orientations of the rotation axis, and the magnitudes of angular momentum for Models G1, G2, and G3.
\label{fig_modelG}}
\end{figure}

\begin{figure}
\epsscale{0.90}
\plotone{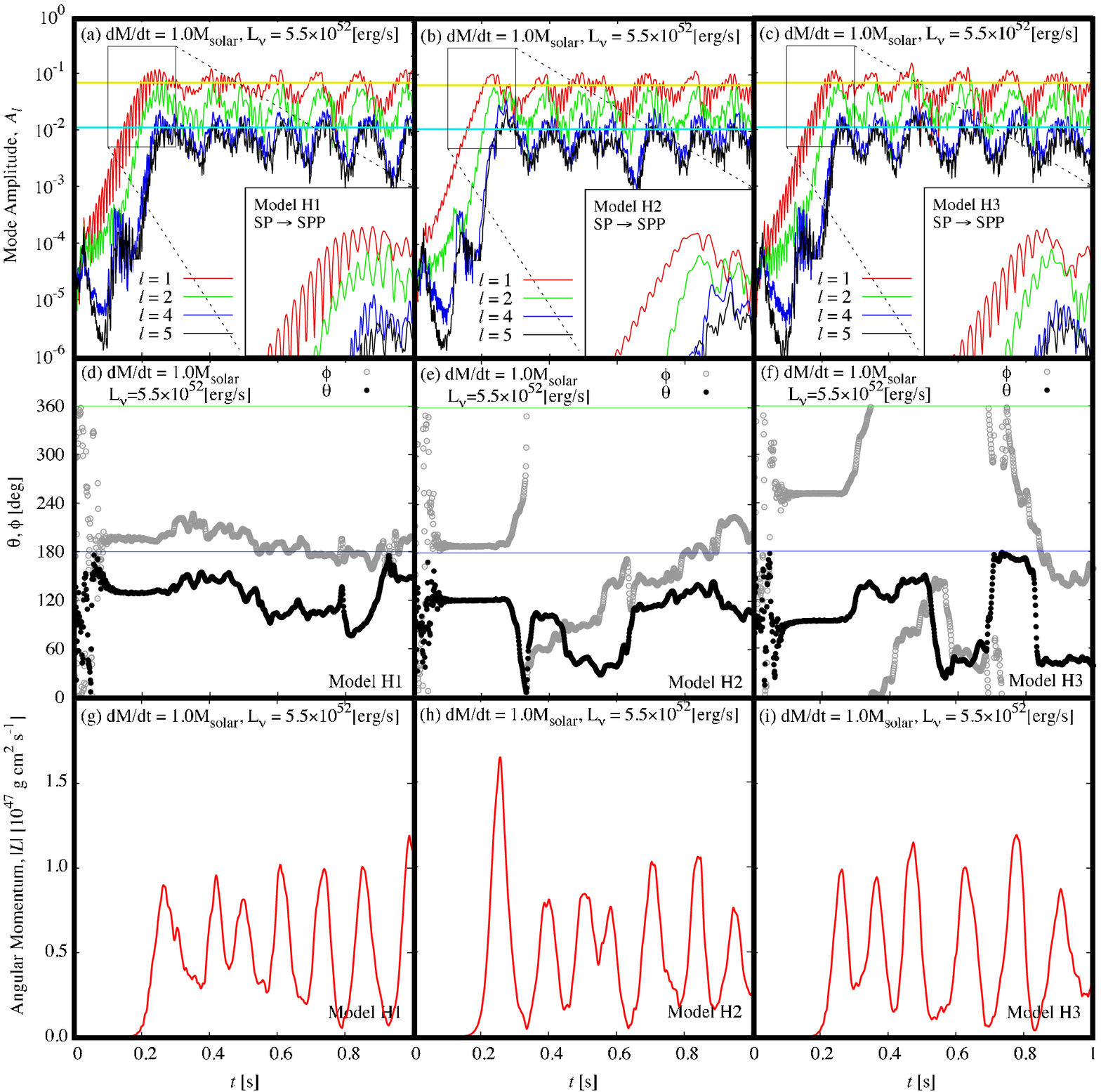}
\caption{
Time evolutions of the normalized mode amplitudes, the orientations of the rotation axis, and the magnitudes of angular momentum for Models H1, H2, and H3.
\label{fig_modelH}}
\end{figure}

\begin{figure}
\epsscale{0.90}
\plotone{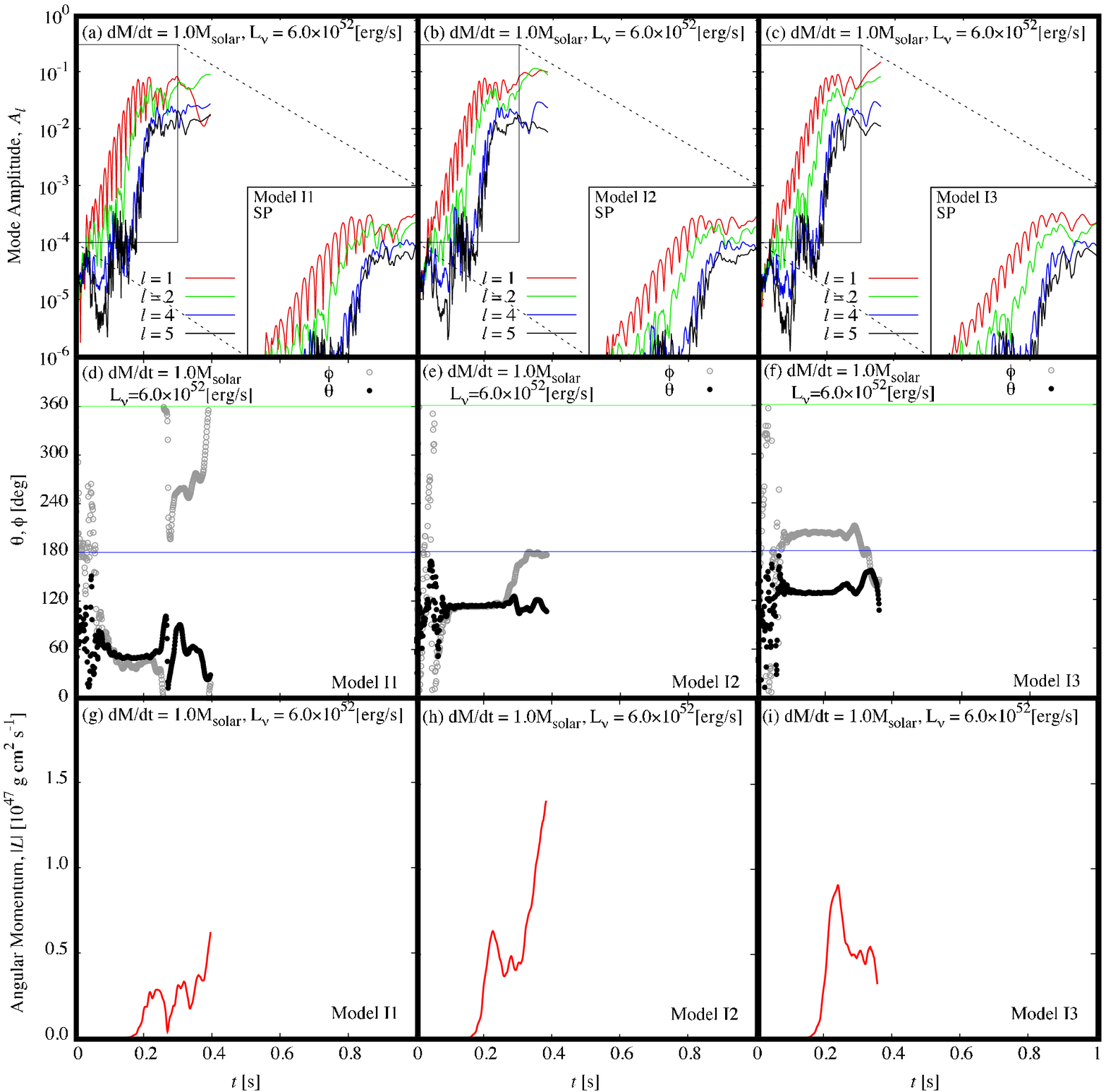}
\caption{
Time evolutions of the normalized mode amplitudes, the orientations of the rotation axis, and the magnitudes of angular momentum for Models I1, I2, and I3.
\label{fig_modelI}}
\end{figure}

In this section, we discuss the reproducibility of the flow patterns, imposing different random perturbations on the initial flows.
Since the post-shock flow is highly stochastic, it is non-trivial whether the flow patterns we have discussed are robust or not.
Three more realizations are presented for each model here.
Taking model A as an example, we name these additional models as A1, A2 and A3, while the original model is referred to as A0.
Figures~\ref{fig_modelA}$-$\ref{fig_modelI} show the results of these additional realizations for Models A$-$I.
In the figures, the time evolutions of the mode amplitudes, the orientation and magnitudes of angular momentum are displayed in the upper, middle and lower panels, respectively.
The different realizations reproduce the same flow patterns except for a few cases, which will be discussed in the following.

Figure~\ref{fig_modelA} indicates that there are two flow patterns, SL and SP, which might be possible, in model A.
In fact, Model A1 has essentially the same features as Model A0: the oscillatory exponential growths in the linear and semi-nonlinear phases and the higher saturation amplitudes of $l=1, 2$ modes (Fig.~\ref{fig_mode}a and \ref{fig_modelA}a), the unstable behavior of the rotational axis (Fig.~\ref{fig_direc}a and \ref{fig_modelA}d), and the lower magnitude of angular momentum (Fig.~\ref{fig_angmom}a and \ref{fig_modelA}g).
Model A1 is hence classified into SL.
We find, on the other hand, that Model A3 has clear SP features:
the gradual change of the rotational axis (Fig.~\ref{fig_modelA}f) and the higher magnitude of angular momentum (Fig.~\ref{fig_modelA}i).
Model A2 occupies a place in between:
the sloshing motions in the semi-nonlinear and nonlinear phases until around 300ms turns into the spiral motions after that (Fig.~\ref{fig_modelA}b, \ref{fig_modelA}e and \ref{fig_modelA}h).
It is thus evident that small differences in the initial perturbations leads to the bifurcation of the flow pattern in the semi-nonlinear and nonlinear phases.
It seems that SP is a little bit more stable than SL.
In fact, we add more 7 realizations for this model and find that SL appears in 4 out of the total 11 realizations.
It is also noted that the direction of the sloshing motion tends to be aligned with the $z$ axis,
which may be a numerical artifact of the spherical coordinates.

Figure~\ref{fig_modelH} shows different realizations of Model H, in which we consistently observe that SP in the semi-nonlinear phase is changed to SPP in the nonlinear phase.
The stability of the rotational axis is different among H0, H1, H2 and H3, though.
The rotational axis is almost fixed in H1 (Fig.~\ref{fig_modelH}d), whereas its orientation is mildly varied for H2 (Fig.~\ref{fig_modelH}e), and drastically changed for H0 and H3 (Fig.~\ref{fig_direc}h and \ref{fig_modelH}f).
The time evolutions of the mode amplitude and angular momentum indicate that large fluctuations of their values are observed as a common feature of these models, local minimum values of angular momentum, on the other hand, tend to be larger for Models H1 and H2 (Fig.\ref{fig_modelH}g and \ref{fig_modelH}h) than for Models H0 and H3 (Fig.\ref{fig_angmom}b and \ref{fig_modelH}i), which corresponds to
the less-frequent changes of the rotational axis for Models H1 and H2.
As mentioned in the earlier sections, the orientation of the rotational axis changes when the rotation almost stops.
The local minimum values of angular momentum are rather stochastic.

For the rest of models, Figure~\ref{fig_modelB}, \ref{fig_modelC}, \ref{fig_modelD}, \ref{fig_modelE}, \ref{fig_modelF}, \ref{fig_modelG} and \ref{fig_modelI} show that the qualitative features of the flow patterns are essentially unchanged among different realizations.
As an example, we take a look at Model G in Figure \ref{fig_modelG}.
SP always emerges with an almost fixed rotation axis both in the semi-nonlinear and nonlinear phases (Fig.~\ref{fig_direc}g, \ref{fig_modelG}d, \ref{fig_modelG}e and \ref{fig_modelG}f).
The angular momentum has similar magnitudes (Fig.~\ref{fig_angmom}g, \ref{fig_modelG}g, \ref{fig_modelG}h and \ref{fig_modelG}i).
The orientation of the rotational axis is random, on the other hands, which is a consequence of the intrinsically stochastic nature of the post-shock flow.
The similar things can be said irrespective of their particular flow patterns for other models (B, C, D, E, F and I): the flow pattern itself is identical among different realizations;
some properties such as the direction of SL/SP, the instantaneous shock geometry of BB, have stochasticity. 
In the explosion models, in particular, the time of the shock wave arrival at the outer boundary (Table~\ref{tbl_param}) and the magnitude of angular momentum behind the shock wave at that time (Fig.~\ref{fig_modelI}g, \ref{fig_modelI}h and \ref{fig_modelI}i) are quite different from realization to realization.

\subsection{Numerical Resolution \label{sec_gcheck}}
For the study of turbulent flows, high enough numerical resolutions are critically important.
In this section, we discuss the dependence of the results on the space and time resolutions.
So far we have used a mesh having $300\times30\times60$ cells on the spherical coordinate system, and adopted the Courant number of 0.5. 
Because of the limitation of numerical resources available to us we add two more models with higher resolutions:
one with $300\times50\times100$ mesh points, and the other model with $300\times60\times120$ mesh points; the Courant number is set to 0.2 in the former model and 0.1 for the latter.
We refer to these two cases by attaching 4 and 5, respectively, to each model name.

\begin{figure}
\epsscale{0.90}
\plotone{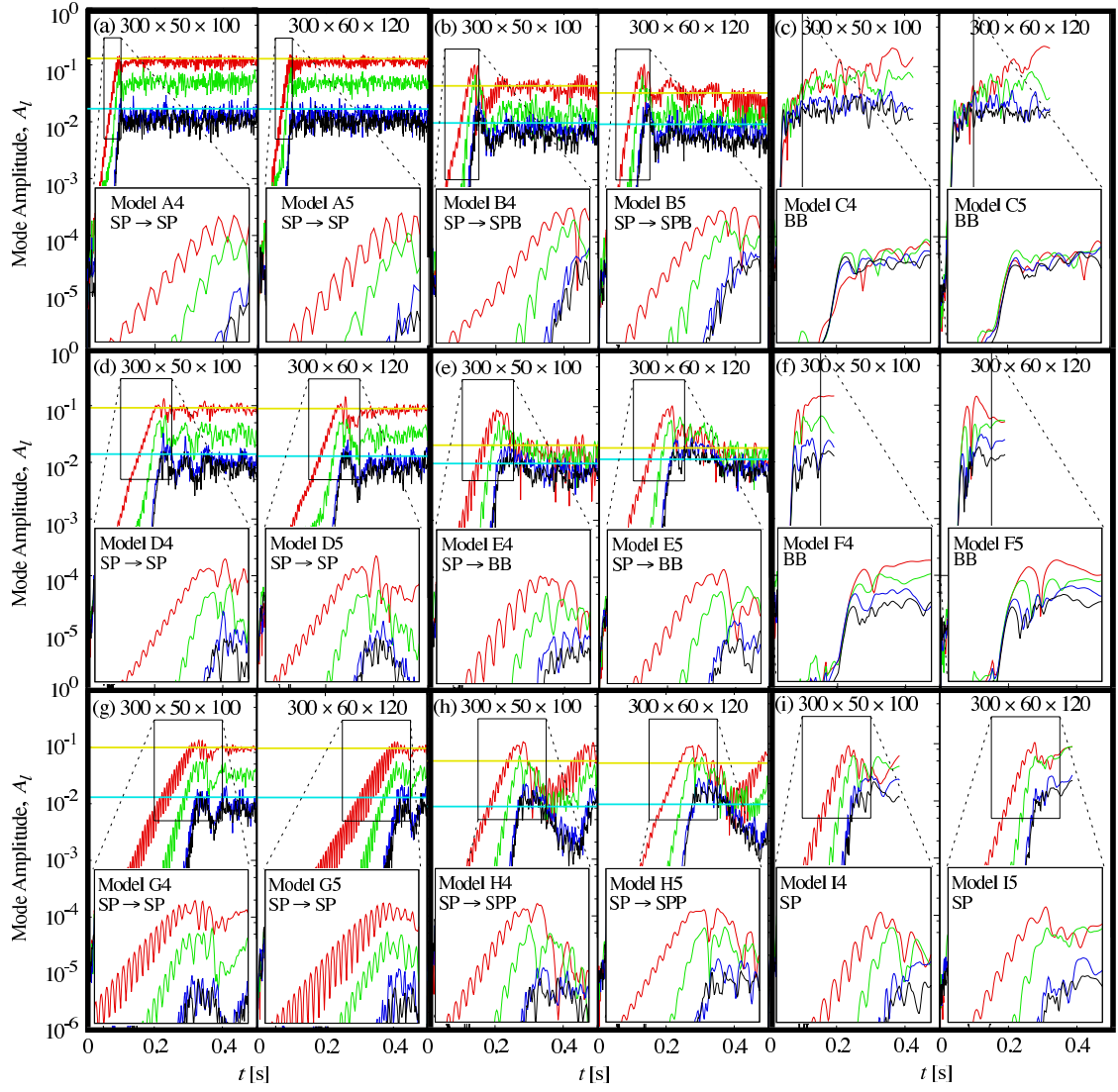}
\caption{
Time evolutions of the normalized mode amplitudes.
From (a) through (i), the left and right panels show the results of 300$\times$50$\times$100 and 30$\times$60$\times$120 mesh points, respectively.
The insets are the zoom-ups of the indicated portions.
\label{fig_mode_check}}
\end{figure}

\begin{figure}
\epsscale{0.90}
\plotone{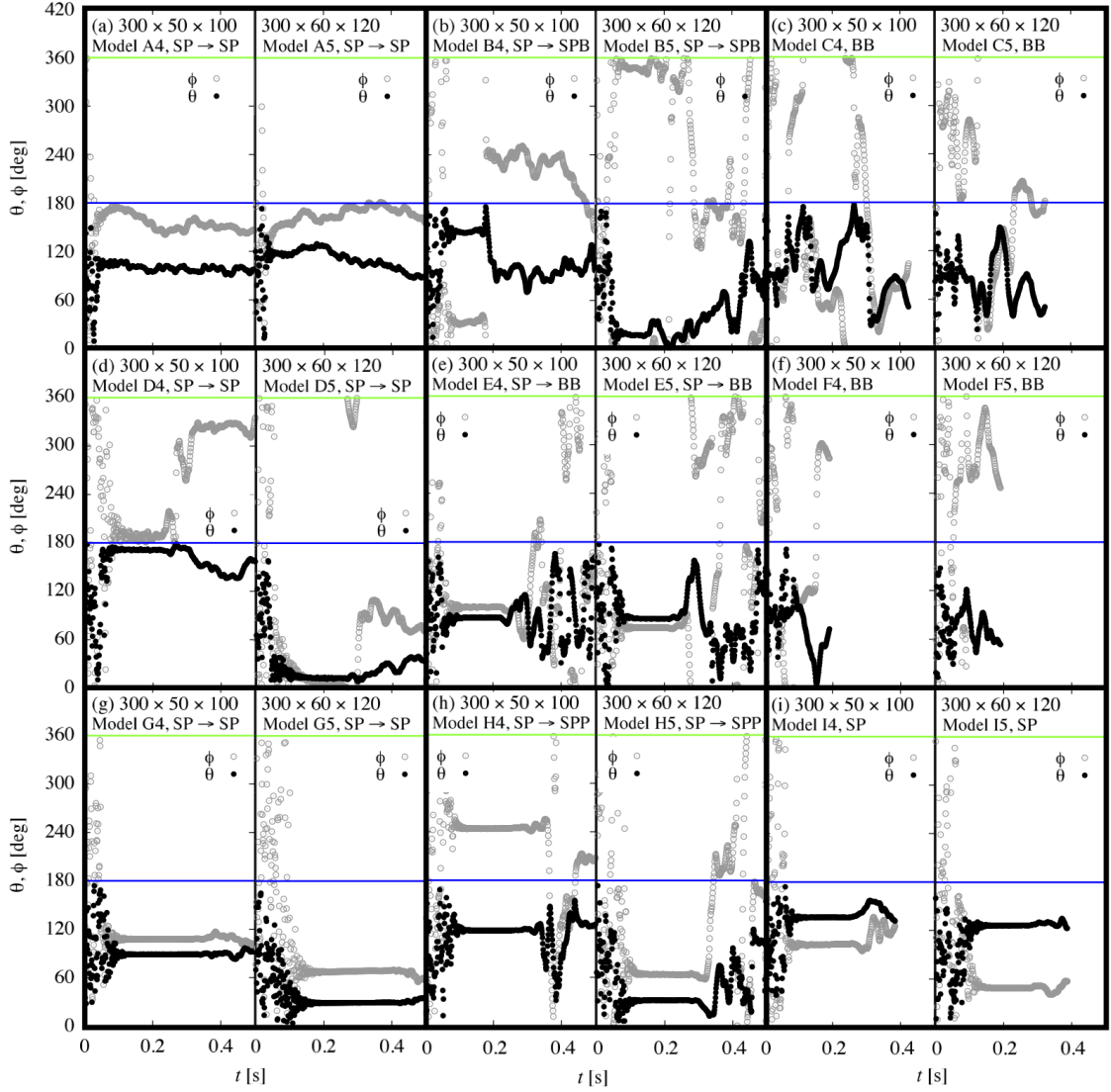}
\caption{
Time evolutions of the orientations of the rotation axis.
From (a) through (i), the left and right panels show the results of 30$\times$50$\times$100 and 30$\times$60$\times$120 mesh points, respectively.
The blue and green lines indicate the positions of 180 and 360 degrees, respectively.
\label{fig_direc_check}}
\end{figure}

\begin{figure}
\epsscale{0.90}
\plotone{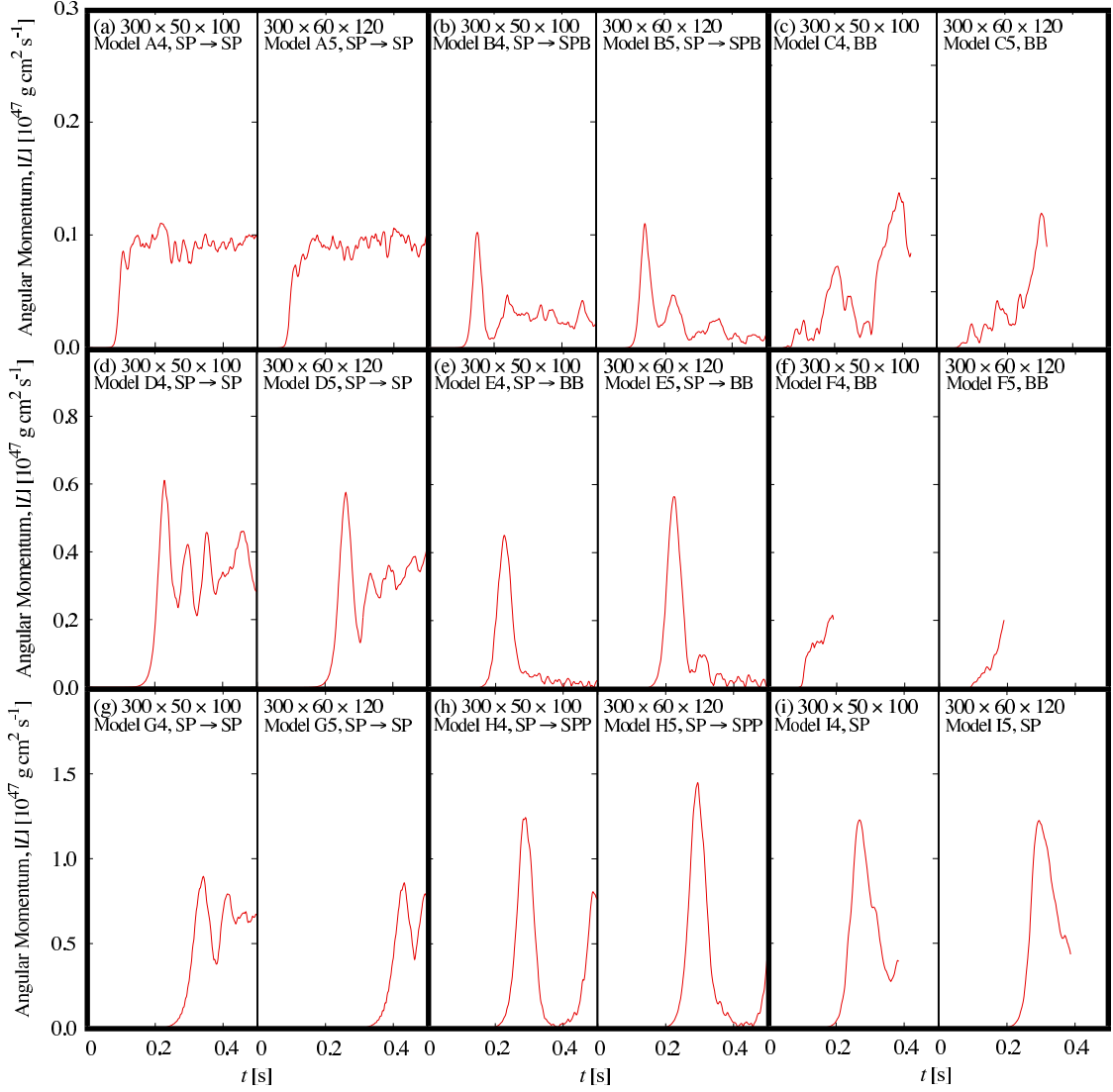}
\caption{
Time evolutions of the magnitudes of angular momentum integrated over the post-shock flows.
From (a) through (i), the left and right panels show the results of 30$\times$50$\times$100 and 30$\times$60$\times$120 mesh points, respectively.
\label{fig_angmom_check}}
\end{figure}

The flow patterns, the average $\chi$ parameter $\bar{\chi}$ and the ratio of average mode amplitudes $\bar{A}_{1,2}/\bar{A}_{3,4}$ are listed in Table \ref{tbl_kai_check}.
We found that all the flow patterns obtained with the higher resolution are the same as the ones with the normal resolution both in the semi-nonlinear and nonlinear phases.
The values of $\bar{\chi}$ and $\bar{A}_{1,2}/\bar{A}_{3,4}$ for SL, SP and SPP are also consistent.
In Figures \ref{fig_mode_check}, \ref{fig_direc_check}, \ref{fig_angmom_check}, we show the time evolutions of the mode amplitudes, orientations of the rotation axis, and magnitudes of angular momentum, respectively.
Their qualitative features are unchanged with the high resolutions except for Model A, in which
the flow pattern appearing in the higher resolution computations, Models A4 and A5 is not SL but always SP like in A3.

There are some differences between the normal and high resolutions, however.
One is the value of $\bar{A}_{1,2}/\bar{A}_{4,5}$ in the buoyant-bubble formations.
In fact, as the number of angular mesh points increases, the absolute values of the ratio $\bar{A}_{1,2}/\bar{A}_{4,5}$ decrease only in SPB and BB, as shown in Table \ref{tbl_kai_check}.
The angle-averaged $\chi$ parameters $\bar{\chi}$ also get larger in BB, which may suggest that
the parasitic instability might be better computed in the higher resolution models (\cite{guilet10}).
In fact, the capture of small structures generated by Kelvin-Helmholtz and/or Rayleigh-Taylor instabilities will require high spatial resolutions.
This is consistent with the just-mentioned fact that the ratio $\bar{A}_{1,2}/\bar{A}_{4,5}$ tends to decrease as the angular mesh points are increased.
Notes however, that this will pose no problem in the identification of the flow pattern.
In fact, the linear analysis for $\dot{M}=1.0M_\odot$ demonstrated that modes with $l\sim 6$ are the most unstable (\cite{yamasaki07}),
which are still enough to be resolved even with the ordinary resolution of 300$\times$30$\times$60 mesh.
Although this is based on the linear analysis, the dominant unstable modes will not be much different for the mean flows.

Another difference we find in the high-resolution simulations is the onset time of the semi-nonlinear phase which is defined in this paper.
The period, in which specific modes start to dominate over other ones after all the modes seeded by the initial random perturbation, grow exponentially in the linear phase.
It is found in some models that the onset of the semi-nonlinear phase is delayed in the high-resolution simulations.
The nonlinear phase that follows the semi-nonlinear phase also commences later as is clearly shown in Fig.~\ref{fig_mode_check}d, \ref{fig_mode_check}g and \ref{fig_mode_check}i.
The end of the linear phase is most easily recognized SP in Fig.~\ref{fig_direc_check}a, \ref{fig_direc_check}b, \ref{fig_direc_check}d, \ref{fig_direc_check}e, \ref{fig_direc_check}g, \ref{fig_direc_check}h and \ref{fig_direc_check}i.
In fact, the time, at which $\theta$ and $\phi$ start to be constant, marks the beginning of the semi-nonlinear phase.
It is also found that the delay of the semi-nonlinear phase is longer for the models with high mass accretion rate.
Note, however, that this may be simply due to the fact that, when the resolution is increased, the shares of large scale modes in the initial cell-by-cell random perturbations become smaller.
Nevertheless, the dominant flow pattern is insensitive to these differences. 
The growth rate of the dominant mode in the semi-nonlinear phase is unaffected, either, as seen for SP in Models A, B, D, E, H, and I, although it is slightly lower in Model G5 than in G0$-$G4.

It is admitted true that the tests presented here are not sufficient to prove the numerical convergence.
It is not easy to increase the resolution further, however, due to the numerical cost.
We should emphasize that in this paper, we are interested only in relatively large and coherent features such as global shock oscillations and formations of large buoyant bubbles.
It is a future task to understand how these coherent structures are formed from a mixture of various modes through nonlinear interactions.

\section{SUMMARIES AND DISCUSSIONS}

Various flow patterns that emerge after the shock stagnation in core-collapse supernovae were investigated in this paper.
We conducted a parametric study, performing three dimensional simulations of post-shock flows in the iron core with the light-bulb approximation for neutrino transfer.
The time evolutions of various mode that are superimposed to represent the shock deformation and those of the orientation and magnitude of the angular momentum integrated over the post-shock flow were used to determine the flow pattern objectively. 
Varying the mass accretion rate and neutrino luminosity systematically, we examined six non-explosion and three explosion models in this study.
The results were essentially classified into three flow patterns: sloshing motion (SL), spiral motion (SP), formation of multiple high-entropy bubbles (BB).
We found in addition two 
intermediate
patterns: spiral motion with buoyant-bubble formation (SPB), and spiral motion with pulsating change of rotational velocities (SPP).
We found that SL and SP occur for high mass accretion rates and low neutrino luminosities, whereas BB appears in the opposite regime with low accretion rates and high neutrino luminosities.
Moreover, it was shown that the $\chi$ parameter originally proposed by \citet{foglizzo07} based on their linear analysis is also useful even in the full-blown turbulence with some possible modifications of the critical value to identify BB.
The findings are consistent with the trends inferred in the preceding works \citep[][]{hanke13, burrows12}.

We divided the growth of the instability into three stages: linear, semi-nonlinear and nonlinear phases, and studied the latter two separately. 
In the semi-nonlinear phases, no 
intermediate
case was observed and only one of the basic three flow patterns was found.
In the case of SL and SP, low $l$ modes grow exponentially with periodic oscillations superimposed.
For BB, on the other hand, we see many modes grow just exponentially.
It is generally thought that SL and SP are associated with the standing accretion shock instability (SASI) which is induced by advective-acoustic cycles, whereas multiple high-entropy bubbles are formed in the neutrino-driven convection, which is caused by a negative entropy gradient.
Although all the initial models studied in this paper have a negative entropy gradient in the gain region, BB appears only when the $\chi$ parameter is larger than a certain value, the fact consistent with the prediction based on linear analysis.
Note, however, that the critical value seems to be slightly larger $\sim 4$ than the originally proposed value $\sim 3$.

In the nonlinear phase, various flow patterns including the 
intermediate
patterns (SPP and SPB) were observed.
In most cases, the dominant flow pattern in the semi-nonlinear phase remains prevalent also in the nonlinear phase. 
In the vicinity of the critical line, SP in the semi-nonlinear phase changes to BB in the nonlinear phase.
The two 
intermediate
patterns were also obtained in the above situation.
One is the spiral motion with pulsating change of rotational velocities (SPP), which occurs for high accretion rates.
The rotational axis rapidly changes its direction when the rotation gets slowed substantially.
The other is the 
intermediate
state between SP and BB that appears for low accretion rates (SPB).
We evaluated the $\chi$ parameter also in this phase, employing the angle-averaged flows.
We found that it is a good measure for BB also in the nonlinear phase.
The critical value seems to be a little bit higher, $\chi \sim 4$ than the canonical values $\chi \sim 3$.
It was demonstrated that it is rather sensitive to the numerical resolution but that the flow pattern itself is robust.

It is natural to ask what mechanism is working behind the formation of the various flow patterns we witnessed so far.
It is not easy to answer this question, though, since complex nonlinear interactions between various modes are at play one way or another.
It will be interesting, however, to mention the nonlinear couplings between SASI and neutrino-driven convections in particular.
\citet{scheck08} demonstrated in their 2D simulations that SASI can trigger convection in the nonlinear phase even for the low neutrino luminosity, for which we do not expect convection normally.
\citet{guilet10} contended that the parasitic instabilities such as Kelvin-Helmholtz and/or Rayleigh-Taylor instabilities grow on top of SASI.
Similar situations might be occurring in our 3D models particularly near the critical neutrino luminosity.
The energy exchange between the large-scale coherent fluid motion by SASI and the multi-scale turbulent fluid motion by neutrino-driven convection have a possibility to occur via nonlinear effects.
When the flow pattern changes from SP in semi-nonlinear phase to SPB/BB in the nonlinear phase, the energy conversion from large-scale coherent motions produced by SASI to multi-scale turbulence induced by neutrino-driven convection might start to be effective toward the end of the semi-nonlinear phase and reach equilibrium in the nonlinear phase.
In this equilibrium the multi-scale turbulent motions may be dominant for BB whereas the large-scale coherent motions would still be competing for SPB.
SPP, on the other hand, might be a state, in which the energy exchange between these two motions occur repeatedly. 
It is emphasized that these are just speculations and detailed analyses should be attempted further but will be deferred to future works.

Finally, changing the initial random perturbation, we investigated different realizations for each model.
It turns out that the flow pattern is a robust feature, the same pattern being almost always reproduced.
There is an exceptional case, however, in which either SL or SP is observed in different realizations.
It is interesting to mention the SWASI experiments by \citet{foglizzo12}, in which  a similar bi-stability was observed.
Judging from the appearance frequency, we infer that SP is slightly more favored in this particular models.
We found some differences among different realizations, which reflect the stochastic nature of the post-shock flows.
Among them are the directions of SL/SP, the frequency of the sudden changes in the orientation of the rotation axis in SPP, the time of shock arrival at the outer boundary in the explosion models, the angular momentum left behind, and so on.

In this study we made a couple of assumptions: constant mass accretion rates, proto-neutron star mass, and neutrino luminosities; the initial condition is spherically symmetric, steady, shocked accretion flows with small random perturbations; neutrino transfer is treated by the light bulb approximation.
The influences of these approximations on our conclusions should be assessed by more realistic simulations.
Note, however, that our results are consistent with other previous works, including some realistic simulations.
At the very end, we mention that it will be interesting to consider the possibility to observationally distinguish the various flow patterns by neutrino and/or gravitational wave signals.
The possible correlation of the flow pattern with the eventual morphology of ejecta may be also worth investigations.
We are actually working in these directions currently.
Last but not least, rotation is another interesting twist, which was entirely ignored in this paper but will be addressed in the forthcoming paper.

\acknowledgments

Numerical computations were performed on the XC30 and the general common use computer system at the center for the Computational Astrophysics, CfCA, the National Astronomical Observatory of Japan, as well as, the Altix UV 1000 at the IFS in Tohoku University and SR16000 at YITP in Kyoto University.
This study was supported by the Grants-in-Aid for the Scientific Research (NoS.~24244036, 24740165), the Grants-in-Aid for the Scientific Research on Innovative Areas, "New Development in Astrophysics through multi messenger observations of gravitational wave sources" (No.~24103006), and the HPCI Strategic Program from the Ministry of Education, Culture, Sports, Science and Technology (MEXT) in Japan.



\clearpage

\begin{deluxetable}{ccccccc}
\tablecaption{Summary of parameters for all models.\label{tbl_param}}
\tablewidth{11cm}
\tablehead{Model & $\dot{M}$\tablenotemark{a} & $L_{\nu}$\tablenotemark{b} & $r_\mathrm{in}$\tablenotemark{c} & $r_\mathrm{out}$\tablenotemark{d} & $t_\mathrm{exp}$\tablenotemark{e}\\
& [$M_\odot$ $s^{-1}$] & [$10^{52}$ erg $s^{-1}$] & [km] & [km] & [ms]
}
\startdata
A & 0.2 & 2.0 & 29 & 581  & -       \\
B & 0.2 & 2.5 & 33 & 655  & -       \\
C & 0.2 & 3.0 & 36 & 712  & 270-352 \\
D & 0.6 & 4.0 & 41 & 822  & -       \\
E & 0.6 & 4.5 & 44 & 872  & -       \\
F & 0.6 & 5.0 & 46 & 920  & 201-208 \\
G & 1.0 & 5.0 & 46 & 919  & -       \\
H & 1.0 & 5.5 & 49 & 965  & -       \\
I & 1.0 & 6.0 & 51 & 1007 & 358-411 \\
\enddata
\tablenotetext{a}{The mass accretion rate.\vspace{-5mm}}
\tablenotetext{b}{The neutrino luminosity.\vspace{-5mm}}
\tablenotetext{c}{The radial position of the inner boundary.\vspace{-5mm}}
\tablenotetext{d}{The radial position of the outer boundary.\vspace{-5mm}}
\tablenotetext{e}{The time for the shock wave to reach $r_\mathrm{out}$.\vspace{-5mm}}
\end{deluxetable}

\begin{deluxetable}{cccccccccc}
\tablecaption{Flow patterns and some parameters in the semi-nonlinear and nonlinear phases.\label{tbl_kai}}
\rotate
\tablewidth{21.5cm}
\tablehead{
Model &
Pattern &
$\chi$\tablenotemark{a} &
$r_{\rm gain}$\tablenotemark{b} &
$r_{\rm sh}$\tablenotemark{c} &
Pattern &
$\bar{\chi}$\tablenotemark{d} &
$\bar{r}_{\rm gain}$\tablenotemark{e} &
$\bar{r}_{\rm sh}$\tablenotemark{f} &
$\bar{A}_{1,2}/\bar{A}_{4,5} $ \tablenotemark{g}\\
&
(semi-nonlinear)&
&
[km] &
[km] &
(nonlinear)&
&
[km] &
[km] &
\\
}
\startdata
A & SL/SP & 1.2 & 52 & 68  & SL/SP & 2.4/2.2 & 49/49 & 107/104 & 9.6/7.3 \\
B & SP    & 2.7 & 56 & 89  & SPB   & 3.8     & 55    & 127    & 5.1     \\
C & BB    & 4.9 & 63 & 112 & -     & -       & -     & -      & -       \\
D & SP    & 1.6 & 71 & 92  & SP    & 2.1     & 69    & 139    & 6.7     \\
E & SP    & 2.5 & 75 & 109 & BB    & 3.6     & 74    & 142    & 3.0     \\
F & BB    & 4.1 & 82 & 132 &  -    & -       & -     & -      & -       \\
G & SP    & 1.4 & 78 & 97  & SP    & 1.6     & 77    & 143    & 7.4     \\
H & SP    & 2.0 & 82 & 110 & SPP   & 2.5     & 81    & 182    & 6.0     \\
I & SP    & 2.9 & 87 & 128 & -     & -       & -     & -      & -       \\
\enddata
\tablenotetext{a}{$\chi$ parameter in the initial flow.\vspace{-5mm}}
\tablenotetext{b}{The gain radius in the initial flow.\vspace{-5mm}}
\tablenotetext{c}{The shock radius in the initial flow.\vspace{-5mm}}
\tablenotetext{d}{$\chi$ parameter in the time- and angle-averaged flow during the nonlinear phase.\vspace{-5mm}}
\tablenotetext{e}{The gain radius in the time- and angle-averaged flow during the nonlinear phase.\vspace{-5mm}}
\tablenotetext{f}{The shock radius in the time- and angle-averaged flow during the nonlinear phase.\vspace{-5mm}}
\tablenotetext{g}{The ratio of the time-averaged mode amplitudes of $A_{1,2}$ to $A_{4,5}$.\vspace{-5mm}}
\tablecomments{
The flow patterns are classified into sloshing motion (SL), spiral motion (SP), multiple buoyant bubbles (BB), spiral motion with buoyant-bubble formation (SPB), and spiral motion with pulsating change of rotational velocities (SPP). For the explosion models, only flow patterns in the semi-nonlinear phase are written in this table.
All values listed above are average ones among different realizations.}
\end{deluxetable}

\clearpage

\begin{deluxetable}{ccccc}
\tablecaption{Flow patterns and physical parameters in the higher resolution.\label{tbl_kai_check}}
\tablewidth{12cm}
\tablehead{
Model &
Pattern &
Pattern &
$\bar{\chi}$ &
$\bar{A}_{1,2}/\bar{A}_{4,5}$ \\
&
(semi-nonlinear)&
(nonlinear) &
&
}
\startdata
A0-3 & SL/SP & SL/SP & 2.3-2.4/2.2 & 9.5-9.6/7.3 \\
A4   & SP    & SP    & 2.4         & 7.2         \\
A5   & SP    & SP    & 2.3         & 7.4         \\
\tableline
B0-3 & SP    & SPB   & 3.7-3.8     & 4.9-5.3     \\
B4   & SP    & SPB   & 3.7         & 4.4         \\
B5   & SP    & SPB   & 3.7         & 3.6         \\
\tableline
D0-3 & SP    & SP    & 1.9-2.3     & 6.6-6.8     \\
D4   & SP    & SP    & 2.0         & 6.8         \\
D5   & SP    & SP    & 1.9         & 6.8         \\
\tableline
E0-3 & SP    & BB    & 3.5-3.6     & 2.8-3.2     \\
E4   & SP    & BB    & 3.8         & 2.1         \\
E5   & SP    & BB    & 4.1         & 1.8         \\
\tableline
G0-3 & SP    & SP    & 1.5-1.8     & 7.4-7.5     \\
G4   & SP    & SP    & 1.8         & 7.4         \\
G5   & SP    & SP    & 1.5         & 7.5         \\
\tableline
H0-3 & SP    & SPP   & 2.4-2.6     & 5.9-6.1     \\
H4   & SP    & SPP   & 2.3         & 6.3         \\
H5   & SP    & SPP   & 2.4         & 5.3         \\
\enddata
\tablecomments{
Models A0-3, A4, and A5 correspond to the results of $300 \times 30 \times 60$, $300 \times 50 \times 100$, and $300 \times 60 \times 120$ mesh points, respectively, and other models are also named in a similar way.
In the normal resolution models, the ranges of $\chi$ parameter taken among different realizations are listed in this table.
The meanings of symbols are referred in Table \ref{tbl_kai}.}
\end{deluxetable}


\clearpage

\clearpage 

\clearpage

\clearpage

\clearpage

\clearpage

\clearpage

\clearpage

\clearpage

\clearpage

\clearpage

\clearpage

\clearpage

\clearpage

\clearpage

\clearpage

\clearpage

\clearpage

\clearpage

\clearpage

\end{document}